\documentclass[reprint,amsmath,amssymb,rmp,aps,superscriptaddress,nofootinbib]{revtex4-2}
\usepackage{graphicx}
\usepackage{bm}
\usepackage[usenames,dvipsnames,table,xcdraw]{xcolor}
\definecolor{goodgreen}{rgb}{0.1,0.5,0}
\definecolor{goodred}{rgb}{0.7,0,0}
\usepackage{amsmath} 
\usepackage[colorlinks,urlcolor=goodgreen,citecolor=blue,linkcolor=goodred]{hyperref}
\usepackage[normalem]{ulem}

\begin{document}

\title{Colloquium: Synthetic quantum matter in non-standard geometries}

\author{Tobias Grass}
\affiliation{Donostia International Physics Center (DIPC), Manuel de Lardizábal 4, 20018 San Sebasti\'an, Spain}
\affiliation{IKERBASQUE, Basque Foundation for Science, Plaza Euskadi 5, 48009 Bilbao, Spain}
\author{Dario Bercioux}
\affiliation{Donostia International Physics Center (DIPC), Manuel de Lardizábal 4, 20018 San Sebasti\'an, Spain}
\affiliation{IKERBASQUE, Basque Foundation for Science, Plaza Euskadi 5, 48009 Bilbao, Spain}
\author{Utso Bhattacharya}
\affiliation{Institute for Theoretical Physics, ETH Zurich, 8093 Zurich, Switzerland}
\author{Maciej Lewenstein}
\affiliation{ICFO-Institut de Ciencies Fotoniques, The Barcelona Institute of Science and Technology, 08860 Castelldefels (Barcelona), Spain}
\affiliation{ICREA, Pg. Lluis Companys 23, 08010 Barcelona, Spain}
\author{Hai Son Nguyen}
\affiliation{Univ Lyon, Ecole Centrale de Lyon, CNRS, INSA Lyon, Universit\'{e} Claude Bernard Lyon 1, CPE Lyon, CNRS, INL, UMR5270, Ecully 69130, France}
\affiliation{Institut Universitaire de France (IUF), F-75231 Paris, France}
\author{Christof Weitenberg}
\affiliation{IQP — Institut f{\"u}r Quantenphysik, Universit{\"a}t Hamburg, Luruper Chaussee 149, 22761 Hamburg, Germany}
\affiliation{The Hamburg Centre for Ultrafast Imaging, Luruper Chaussee 149, 22761 Hamburg, Germany}
\affiliation{Department of Physics, TU Dortmund University, 44227 Dortmund, Germany}

\begin{abstract}
Quantum simulation is making a significant impact on scientific research. The prevailing tendency of the field is to build quantum simulators that get closer to real-world systems of interest, in particular electronic materials. However, progress in the microscopic design also provides an opportunity for an orthogonal research direction: building quantum many-body systems beyond real-world limitations. This colloquium takes this perspective: Concentrating on synthetic quantum matter in non-standard lattice geometries, such as fractal lattices or quasicrystals, higher-dimensional or curved spaces, it aims at providing a fresh introduction to the field of quantum simulation aligned with recent trends across various quantum simulation platforms, including atomic, photonic, and electronic devices. We also shine light on the novel phenomena which arise from these geometries: Condensed matter physicists may appreciate the variety of different localization properties as well as novel topological phases which are offered by such exotic quantum simulators. But also in the search of quantum models for gravity and cosmology, quantum simulators of curved spaces can provide a useful experimental tool.
\end{abstract}

\maketitle
\tableofcontents

\section{Introduction and outline}
As envisioned by R. Feynman in 1982~\cite{Feynman1982}, a quantum simulator is a controllable quantum system that is used to imitate another quantum system of interest. Since Feynman's early inspiration, many quantum simulation platforms have now been developed~\cite{Altman2021}, ranging from atomic to photonic and electronic devices. Common to all these systems is their synthetic nature providing control over the Hamiltonian.
Quantum simulators have been used to study quantum phenomena across the whole energy spectrum, from low-temperature behavior of quantum materials to out-of-equilibrium phases and quantum chemistry to high-energy physics and cosmological models. Importantly, the application of quantum simulators has not remained limited to the mere imitation of existing systems, but an increasing amount of experimental and theoretical efforts are dedicated to also implementing quantum systems without real-world counterparts. Through these efforts, the quantum-mechanical consequences of exotic geometries, as an alternative to the Euclidean three-dimensional space we are used to, become the attainable subject of modern science.

Examples of such exotic geometries include various non-periodic lattices, from prototypical quasiperiodic models in one or two dimensions to fractal lattice structures with a fractal Hausdorff dimension, but also curved lattices, for instance, hyperbolic lattices. In addition, the scheme of synthetic dimension allows to build systems with more than three space-like dimensions, and in this way expose quantum systems to an ``augmented quantum reality''. The motivations for these efforts are manifold: Space-time curvature is a key element of cosmological or quantum gravity models but typically occurs on scales that are hard to probe experimentally. Quantum simulators of curved spaces bring quantum field theory in curved space-time into the lab and thereby provide a tool for analog studies of cosmological models~\cite{Viermann2022}, Unruh effect~\cite{Volovik_2016,Rodriguez-Laguna2017,Kosior2018,Armitage_2018}, or analog black holes~\cite{Garay2000,Nguyen2015,Steinhauer2016}.

Also, on a totally different length scale, in the context of microscopic behavior in condensed matter, synthetically generated settings can provide a fresh perspective on fundamental aspects. Various quantum aspects of matter, both on the single- and the many-body level, strongly depend on the dimensionality and geometry of the space in which matter is embedded. Examples are quantum transport, localization and thermalization behavior, topological transport, and topological classification of quantum systems. For instance, localized 1D or 2D systems with a mobility edge are exclusive to carefully designed quasiperiodic potentials~\cite{Biddle2011,Ganeshan2015,Lueschen2018,Deng2019,An2021}. In synthetic quasicrystals with interactions, the nearly perfect isolation from the environment has enabled the observation of a phenomenon known as many-body localization (MBL)~\cite{Schreiber2015,Lukin2019}, the absence of thermalization. The 4D quantum Hall effect is an exciting example of a topological system that involves topological concepts, known as the second Chern number, beyond the ones relevant to real-world 2D topological matter~\cite{Kraus2013}. Fractal lattices provide a playground to explore topological order without a true bulk~\cite{Brzezinska2018,Pai2019,Iliasov2020,Fremling2020,Biesenthal2022}. Quasicrystals, characterized by symmetries impossible in regular crystals, open the door to novel symmetry-protected topological phases~\cite{Varjas2019,Else2021}.

\begin{table*}
\includegraphics[width=0.99\textwidth]{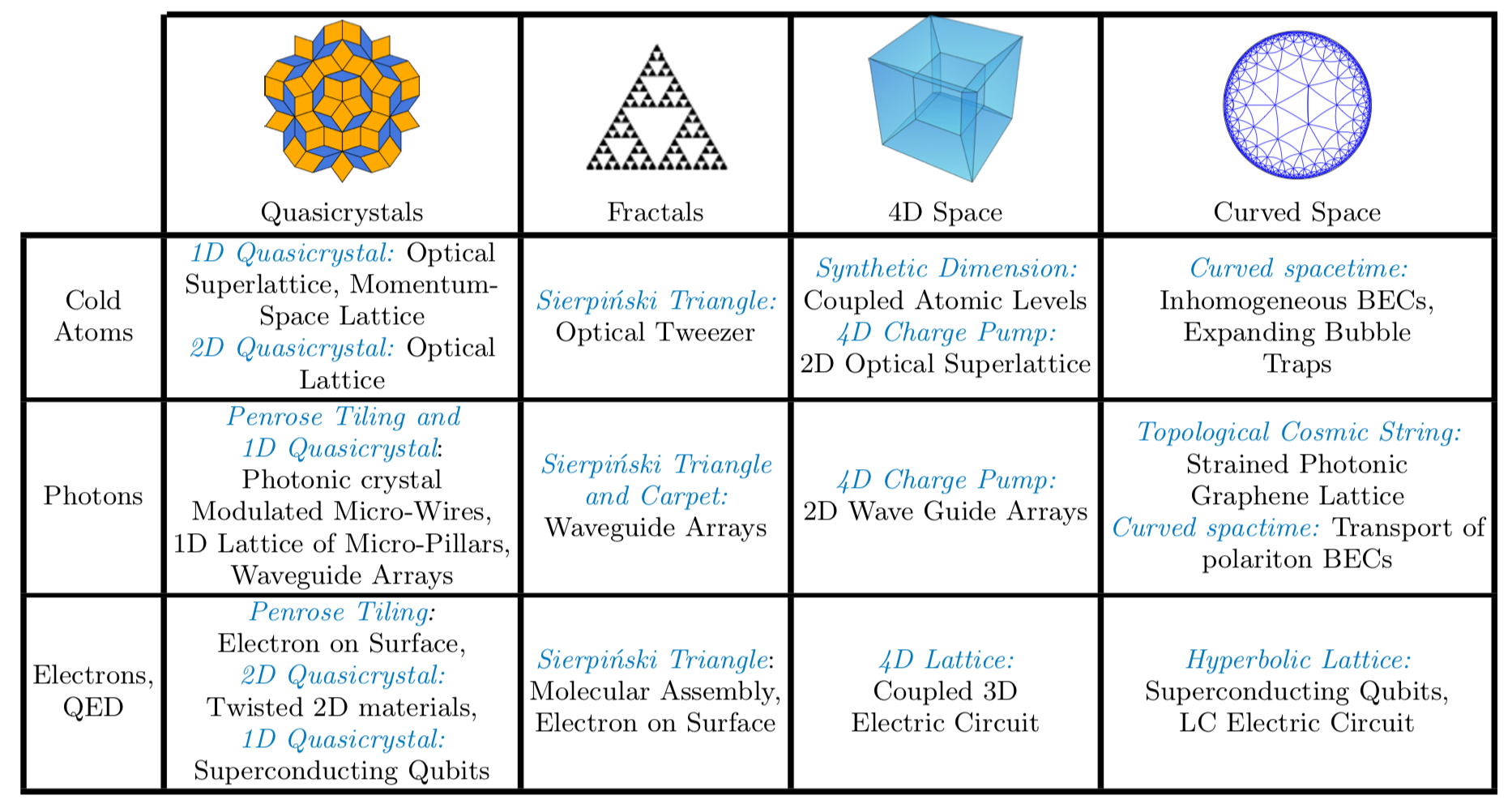}
\centering
    \caption{
    Overview of different exotic geometries and their implementations in different quantum simulation platforms. In blue slanted, we indicate the type of nonstandard geometry, and in black, the method of realization, details are given throughout the text.}
       \label{tab:ov}
\end{table*}

In the present colloquium, we showcase implementations of quantum systems in exotic geometries across different experimental platforms and then discuss the novel physical phenomena that can be addressed in these systems. Specifically, the geometries in the focus of this article are:

\paragraph{Quasicrystals.\label{Intro:quasicrystals}} Quasicrystals are a third kind of solids, between the periodic crystals and the amorphous solids or glasses, which, typically due to a non-negligible amount of disorder, lack translational symmetry.
The structure of quasicrystals follows a deterministic construction rule, which gives rise to long-range order as in ordinary crystals. However, like glasses, they are aperiodic. Physically, their order becomes manifest, for instance, through Bragg peaks in a diffraction experiment, which also has played a crucial role in the discovery of quasicrystals in 1984~\cite{Shechtman1984, Levine1984}, when D. Shechtman and co-workers found aluminum-manganese alloys with tenfold rotational symmetry~\cite{Shechtman1984}. Indeed, some definitions of a quasicrystal demand that it exhibits a rotational symmetry incompatible with periodic structures limited to two-, three-, four-, or sixfold rotational symmetry. According to a more general definition that also admits 1D quasicrystals, this term is simply an abbreviation for quasiperiodic crystal~\cite{Lifshitz2003}.

Shechtman's discovery, awarded the Nobel Prize in Chemistry in 2011, extended the research on aperiodic tilings from mathematically and/or aesthetically grounded studies in the 1970s~\cite{Penrose1974} to physics and material science. 
Nowadays, research on quasicrystals is gaining importance through the rapid progress of synthetic materials, which have opened the door to experimental studies of an even larger variety of quasicrystals, including paradigmatic 1D models. On the tight-binding level, these models are of the form 
%
%
\begin{equation}
H_{\rm tb}= \sum_n \left( -t_n a_{n+1}^\dagger a_n + {\rm h.c.} + V_n a_n^\dagger a_n \right),
\label{AAH}
\end{equation}
%
%
with $a_n^\dagger$ ($a_n$) creation (annihilation) operators on site $n$. The parameters $t_n$ and $V_n$ are neither periodic (as in crystal models) nor random (as in glassy models), but (at least) one of them is obtained from a quasiperiodic sequence, such as the Fibonacci sequence, see~\textcite{Jagannathan2021} for a review. For instance, the diagonal Fibonacci quasicrystal has constant hopping, $t_n=t$, but the potential $V_n$ is a sequence of binary potential values. This sequence is generated iteratively by merging the previous two iterations. Another famous 1D quasicrystal model is the Aubry-Andr\'e (AA) model with
%
%
\begin{equation}
V_n = \lambda \cos(2\pi \alpha n + \phi).
\label{AA}
\end{equation}
%
%
Here, $\alpha$ is irrational, and $\phi$ a random phase. We provide an overview of different realizations of synthetic quasicrystals in the first row of Table \ref{tab:ov}.

\paragraph{Fractals.} Another non-periodic geometry are fractals. Following a deterministic construction scheme, they may appear rather artificial, but fractals are omnipresent in organic nature~\cite{Mandelbrot}. Fractals are characterized by self-similarity: repeating the same patterns on different scales and by a fractal dimension, a non-integer Hausdorff dimension~\cite{Mandelbrot}. The concept of Hausdorff dimension is 
based on box-counting: the method considers a grid with spacing $\epsilon$ of the embedding Euclidean space and counts the number of boxes $N(\epsilon)$ that fit into this space. The Hausdorff dimension $d$ is determined by the scaling of $N(\epsilon)$ in the limit $\epsilon \rightarrow 0$, specifically
\begin{align}
d=- \lim_{\epsilon \rightarrow 0} \frac{\log N(\epsilon)}{\log \epsilon}.
\end{align}
An example of a fractal, the so-called Sierpi{\'n}ski triangle or gasket, is shown on the top of the second column of Table~\ref{tab:ov}. The structure is generated by dividing an equilateral triangle into smaller ones and removing the central smaller triangle. The Hausdorff dimension of this structure is $d=\frac{\log 3}{\log 2}\approx 1.585$. A similar structure, starting from a square instead of a triangle, is known as Sierpi{\'n}ski carpet and has a Hausdorff dimension of $d=\frac{\log 8}{\log 3}\approx 1.893$. The procedure can also be applied to structures embedded in a three-dimensional space, e.g., tetrahedrons. By division into smaller and smaller tetrahedrons combined with the removal of selected tetrahedrons, one ends up at the Sierpi{\'n}ski tetrahedron, which curiously has an integer Hausdorff dimension of 2. Through synthetic systems, fractals have also entered the arena of quantum physics, see Table \ref{tab:ov}, and in particular, the dynamics of quantum particles in fractal space have become the subject of intense research.

\paragraph{Lattices with extra dimension.} Even integer dimensions can be considered quite exotic if the number of dimensions exceeds three. Edwin Abbott's satirical novella \textit{Flatland} from 1884 describes a society of flatlanders living in a 2D world~\cite{Abbott1884}. For them, a 3D sphere that crosses flatland is perceived as a circle that, strangely enough, appears out of nothing and constantly changes its radius until it again shrinks to a point and disappears. The only way to understand the origin of this strange object is by leaving flatland. Modern quantum simulators allow us to leave our 3D perspective on the world, by engineering quantum systems that incorporate other than purely spatial degrees of freedom in a space-like fashion, see Table~\ref{tab:ov}. These experimental systems have opened a door to phenomena occurring only in higher-dimensional spaces, particularly novel types of topological matter~\cite{Price_2022}. 

\paragraph{Curved lattices.} Finally, lattice systems can become exotic if their metric is not the usual Euclidean but exhibits some curvature. Curved spaces have been traditionally investigated in high-energy physics and cosmology. In recent years, there has been a growing interest in realizing tabletop simulations of hyperbolic lattices. The motivations arise from discovering the holographic principle~\cite{Maldacena_1999, Witten_1998} and the characterization of classical and quantum states in spaces with negative curvature. One possible way to investigate curved lattices beyond the Euclidean case is to work on the Poincaré disk. Here, the tessellation of the space is defined via two indices $\{p,q\}$, known as the Schl\"afli symbol. It corresponds to a tessellation of the plane by regular $p$-gons with coordination number $q$~\cite{Coxeter_1973}. Given a $\{p,q\}$, the curvature of the space is defined by
%
%
\begin{align}\label{curvature}
    \mathcal{D}=(p-2)(q-2).
\end{align}
%
%
If $\mathcal{D}<4$, the space is spherical (positive curvature); whereas, if $\mathcal{D}=4$, the space is Euclidean (zero curvature), and finally, if $\mathcal{D}>4$, we have a hyperbolic space (negative curvature)~\cite{Boettcher2022}. Euclidean tesselations are limited to the well-known triangular $\{3,6\}$, square $\{4,4\}$, and hexagonal $\{6,3\}$ lattices. The example of a $\{3,7\}$ hyperbolic lattice is shown in the illustration of Table~\ref{tab:ov} (right column).
Hence, by the choice of connectivity in a lattice $q$, non-Euclidean spaces can be implemented, and in this way, hyperbolic lattices, characterized by negative curvature, have been produced in photonic and electronic quantum simulators, see Table \ref{tab:ov}. Another approach to curved spaces is through an analogy between sound propagation on a background hydrodynamic flow and field propagation in curved space-time, see~\textcite{Unruh1981}. This analogy has allowed for experimental studies of 
gravitational phenomena in flat quantum simulators.

This colloquium reviews how the described variety of exotic geometries is implemented and exploited in contemporary quantum simulators using atomic, photonic, and electronic platforms. Specifically, in the next section ({\bf Implementation}, Sec.~\ref{Part1}), we will review state-of-the-art quantum engineering techniques used to achieve such geometries. In the subsequent section ({\bf Applications}, Sec.~\ref{Part2}), we will discuss the specific implications of exotic geometries on physical system behavior. Here, we will differentiate between three different phenomenological areas: (i) localization phenomena, (ii) topological phenomena, and (iii) analog black holes and cosmology.

\section{Implementation - engineering exotic geometries \label{Part1}}
We will present an overview of state-of-the-art quantum simulation systems/techniques by demonstrating how exotic geometries are engineered in different platforms, covering atomic, photonic, and electronic systems as outlined in Table~\ref{tab:ov}.

\subsection{Atomic systems}

Ultracold atoms are a formidable experimental platform for quantum simulation due to their large tunability of confining potentials and interaction strength and ability to reach strongly-correlated quantum states with long lifetimes. Neutral atoms are routinely cooled to Mikrokelvin temperatures via laser cooling and further cooled to Nanokelvin temperatures via evaporative cooling in magnetic or optical traps. Ultrahigh vacuum conditions allow for thermal insulation from the vacuum chamber walls. Atoms exist as bosons or fermions depending on the total spin of the atomic isotope, and at Nanokelvin temperatures, they form Bose-Einstein condensates (BEC) or degenerate Fermi gases, respectively.

Optical lattices formed by the interference of laser beams then provide a means to study solid-state physics in a system with 10,000 times larger lattice spacing than electronic materials. This makes the optical imaging of each particle possible. From the resulting snapshots of many-body wave functions, arbitrary correlation functions can be extracted, extending our understanding of what observables are experimentally accessible. 

Atoms have internal hyperfine states that can equally be controlled and harnessed as an effective spin. This allows to study quantum magnetism or to engineer spin-orbit coupling via two-photon Raman transitions. Some atoms also have long-lived optically excited states that can be employed as orbital degrees of freedom accessed by single-photon transitions. 

An important technique for further tuning synthetic systems is Floquet engineering~\cite{Bukov2015}. The Floquet theorem states that a periodically-driven system has an effective description via a static Floquet Hamiltonian when probed stroboscopically. This effective Hamiltonian can have new desired properties, such as stable trapping of a rotating saddle point used in ion traps or artificial gauge fields for neutral particles such as cold atoms or photons~\cite{Eckardt2017}. Remarkably, Floquet driving is also a way to go beyond systems with a static counterpart, as in anomalous Floquet topological insulators that break the bulk-boundary correspondence of static systems~\cite{Rudner2013}. Floquet engineering is used in all platforms considered here, i.e., ultracold atoms~\cite{Eckardt2017, Weitenberg2021}, photonic systems~\cite{Ozawa2019} and solid-state systems~\cite{Oka2019,Rudner2020}. 

With this large toolkit available, ultracold atoms have been used to study diverse effects of solid-state physics, high-energy physics, and quantum systems without counterparts in the real world, as further discussed in the following.

\begin{figure}
    \includegraphics[width=0.49\textwidth]{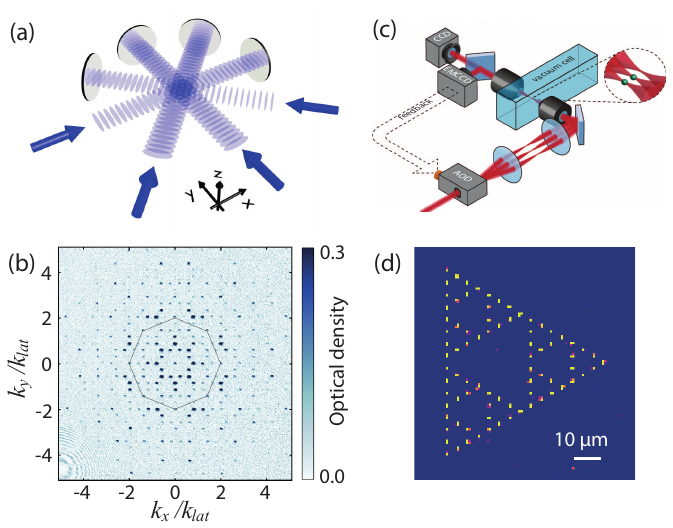}
    \caption{\textbf{Implementation of quasicrystals and fractals with ultracold atoms.} (a) An eightfold quasicrystal lattice can be created by superposing four 1D optical lattices formed by interfering retro-reflected laser beams. (b) Absorption image of the momentum space after pulsing on the lattice for 6 $\mu$s. During the Kapitza-Dirac diffraction dynamics, the atoms populate successively with higher diffraction orders~\cite{Viebahn2019}. Here, $k_{\rm lat}$ is the wavevector of a single lattice beam. (c) Optical tweezers are created by focusing down laser light using high-resolution objectives, and their position can be dynamically arranged by AOD or SLMs~\cite{Endres2016}. (d) Single-shot image of a defect-free atom array with a Sierpi{\'n}ski gasket geometry~\cite{Tian2023}.
    \label{fig:implementation-quasicrystals-fractal}}
\end{figure}

\begin{figure}
    \includegraphics[width=0.49\textwidth]{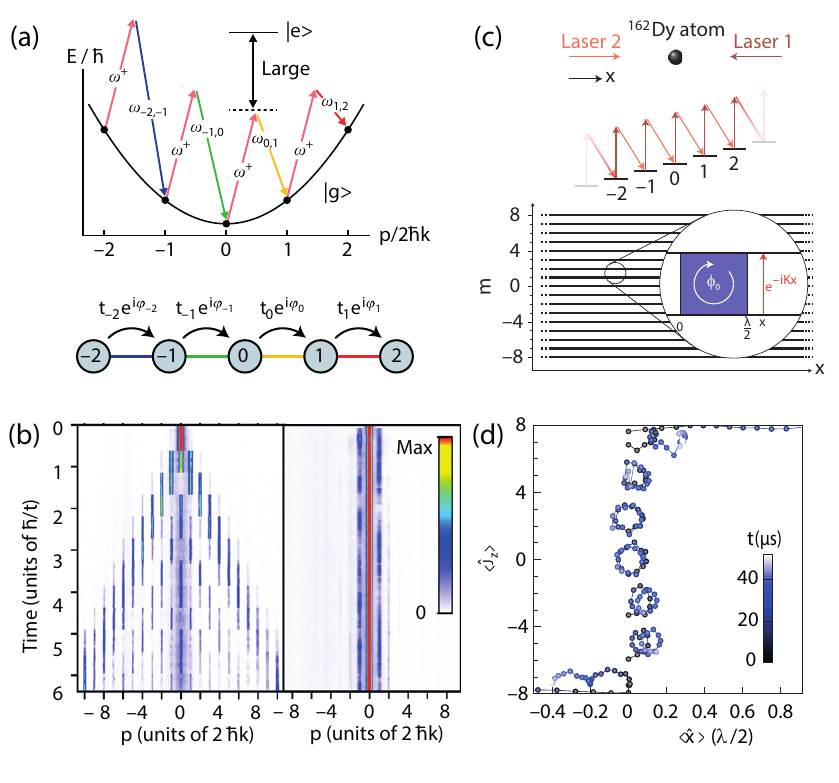}
    \caption{\textbf{Implementation of artificial dimensions with ultracold atoms.} (a) A momentum-space lattice is formed by coupling discrete momentum modes by individually tuned Bragg transitions  driven by laser beams with wavevectors $k$ at different frequencies $\omega$ detuned from the excited state $|\rm e\rangle$. A detuning from the two-photon resonance yields an on-site energy. (b) The resulting dynamics starting from a BEC at zero momentum features a quantum walk for a homogeneous system (left) and localization for pseudo-random site energies (right)~\cite{An2017}. (c) A synthetic dimension is formed by the internal states of an atom coupled by optical Raman transitions. (d) This setting allows adding Peierls phases to create a quantum Hall system, where the cyclotron and skipping orbits can be directly observed~\cite{Chalopin2020}. $\langle J_z \rangle$ measures the spin projection with quantum number $m$ shown in (c).
    \label{fig:implementation-artificial-dimension}}
\end{figure}

\subsubsection{Quasicrystals and fractals}

\paragraph{Quasicrystal optical lattices.}
Optical lattices have been a primary tool for quantum simulation with cold atoms, including exotic geometries~\cite{Windpassinger2013}. Very elegantly, optical lattices can also produce quasicrystalline potentials by an appropriate choice of laser beams forming the lattice. The AA model in Eq.~\eqref{AA}, a 1D lattice with quasi-random on-site energies, was realized by adding a second weaker 1D optical lattice of incommensurate spatial period due to an incommensurate wavelength~\cite{Roati2008,Rajagopal2019}. This approach can be used as an alternative to laser speckle disorder~\cite{Billy2008}, but as will be discussed in
Sec.~\ref{Sec:loc} also allows for fine-tuned localization properties, such as 1D mobility edges. In 2D, fivefold quasicrystal potentials would be formed by the interference of five lattice beams~\cite{Sanchez-Palencia2005}, and a challenge would be to control the relative phases of the laser beams~\cite{Corcovilos2019}. This technical challenge can be overcome using the multi-frequency lattice scheme leading to a pairwise interference of the beams at different frequencies with well-controlled relative phases~\cite{Kosch2022}. An eightfold quasicrystal potential was realized by superposing four 1D optical lattices at slightly different frequencies such that they do not interfere with each other~\cite{Viebahn2019,Sbroscia2020} | see Figs.~\ref{fig:implementation-quasicrystals-fractal}a,b. These experiments have also triggered new theoretical interest in many-body physics in quasicrystal potentials and new descriptions such as Hubbard models for quasicrystalline potentials~\cite{Johnstone2019,Gautier2021,Gottlob2023,Zhu2023}.

\paragraph{Optical tweezers.}
Atomic systems can also be assembled one by one via optical tweezers | see Fig.~\ref{fig:implementation-quasicrystals-fractal}c. These are tightly focused optical dipole traps directly loaded from a magneto-optical trap, where the collisional-blockade mechanism ensures occupation with no more than one atom~\cite{Schlosser2001}. Subsequent detection and rearrangement of the traps then allows for the creation of defect-free arrays~\cite{Endres2016,Barredo2016}, which are now widely used as starting point for quantum computing protocols and quantum simulation~\cite{Semeghini2021}. This approach allows creating arbitrary geometries including Fibonacci quasicrystals~\cite{Wang2020npj} and fractal structures~\cite{Tian2023} | see Fig.~\ref{fig:implementation-quasicrystals-fractal}d. Due to the micrometer-spacing of the tweezers, significant tunnel coupling between them was only reached for the lightest atom lithium~\cite{Spar2022}, but Rydberg excitations can hop between the tweezers and realize relevant quantum many-body models.

\paragraph{Momentum-space lattices.}
As an alternative to optical lattices and optical tweezers, one can also use a mapping between real-space lattices and synthetic lattices composed of discrete internal~\cite{Boada2012} or external~\cite{An2017,Price2017} states. The external states can be discrete momentum states coupled by two-photon Bragg transitions of appropriate frequencies | see Figs.~\ref{fig:implementation-artificial-dimension}a,b. Synthetic lattices offer the great advantage of microscopic control over all system parameters, such that local gauge fields, tailored disorder, or quasiperiodic potentials can be engineered~\cite{An2018PRX,An2021,Wang2022}. Extensions to 2D momentum-space lattices have been discussed~\cite{Agrawal2023}. 

\subsubsection{Lattices with synthetic dimension}
The synthetic lattice made up of Raman-coupled internal states can be combined with a real-space lattice, such that the internal states provide an additional compact dimension to the system~\cite{Boada2012}. This approach facilitates imprinting a Peierls phase on the coupling element within this synthetic dimension, giving rise to an artificial gauge field. Furthermore, the synthetic dimension has a natural sharp edge. Exploiting these features, several groups have realized a synthetic quantum Hall stripe and directly observed the skipping orbits at the system's edge~\cite{Mancini2015,Stuhl2015,Chalopin2020}.
Due to the quadratic Zeeman splitting between internal atomic levels labeled by the magnetic quantum number $m$, it is also possible to spectrally select different Raman transitions, and depending on the polarisation of the laser beams, transitions can either couple states with $\Delta m=1$ or $\Delta m=2$. This feature has been proposed to realize periodic boundary conditions or lattices with nontrivial connectivity such as tori or M\"obius strips~\cite{Boada2015,Anisimovas2016}. A synthetic topological Hall cylinder was recently realized using four internal states of $^{87}$Rb atoms coupled with Raman beams and microwave fields~\cite{Li2022PRXQ}.
Finally, using the large spin $J=8$ of $^{162}$Dy atoms with 17 magnetic sublevels, a 4D system was recently built by combining two real dimensions with two synthetic dimensions in cylindrical geometry given by the magnetic projection $m$ and its remainder $n = m~(\text{mod}~3)$, respectively~\cite{Bouhiron2022}. 
A very short synthetic dimension of just two spin states can be used to emulate a bilayer system, with nontrivial hopping between the layers induced by Raman transitions. This can realize tunable graphene-Haldane bilayers~\cite{Cheng2019} or systems analogous to twisted bilayer graphene featuring flat bands~\cite{Salamon2020}.

\subsubsection{Curved space} 
An optical lattice may also mimic curved space if the hopping amplitudes are modified by the coupling to a gauge field that carries information on the effective curvature~\cite{Boada2011}. The spatially dependent complex hopping amplitudes can be engineered, e.g., by tailoring the profile of the laser beams that drive the laser-assisted hopping as Raman transitions between the two sublattices of a spin-dependent optical lattice. It is predicted that this allows the simulation of the motion of massless fermions in a Rindler metric.

While inhomogeneous tunnel elements have not been experimentally realized so far, there are experiments in bulk systems utilizing inhomogeneous interactions instead. In~\textcite{Viermann2022}, inhomogeneous interaction energy results from an inhomogeneous density distribution in a trapped system and can be used to engineer a metric. In~\textcite{DiCarli2020}, an inhomogeneous interaction strength was engineered using a magnetic Feshbach resonance and a magnetic field gradient. 

A different approach to simulating curved spacetime with ultracold atoms utilized the equivalence of an accelerated frame to a parametric modulation of the interaction strength of an atomic BEC~\cite{Hu2019}. The protocol leads to matter-wave jets, which share their characteristics with the Unruh radiation, namely long-range coherence combined with local thermal distribution.

An exciting way of studying an expanding universe is to work with BECs in supersonically expanding toroidal traps~\cite{Bhardwaj2021}. This was realized with 1D ring traps with variable radius created via a digital micromirror device~\cite{Eckel2018,Banik2022}, finding a redshift of long-wavelength excitations as in an expanding universe as well as the production of solitons and vortices. A 2D version, a bubble trap forming the surface of a sphere~\cite{Garraway2016}, was recently realized via radiofrequency dressing of a magnetic trap using the microgravity conditions on the international space station~\cite{Carollo2022}. The experiment studied the thermodynamics of ultracold bosonic atoms, demonstrating substantial cooling upon increasing the size of the sphere, which links to expansion physics analogue to the expanding universe. Such 2D bubble traps are also relevant for the study of space-curvature effects on quantum many-body physics, e.g., on superfluidity and the dynamics of vortices and solitons~\cite{Caracanhas2022,Tononi_2022,Lundblad_2023,Tononi2024PR,Tononi2024AVS}.

\subsubsection{Interactions} 

A central feature of ultracold atoms is the tunability of the interactions. External fields can enforce resonances between different atomic levels, strongly affecting the atomic scattering length. In particular, magnetic Feshbach resonances have become a standard tool to tune the contact interaction strength continuously~\cite{Chin2010}. In a quantum simulator, this allows to connect the (usually) understood non-interacting regime to the computationally complex interacting regime. 

Next to these contact interactions, there are efforts to engineer long-range interactions, e.g., by using dipolar atoms, molecules, optical cavities, or Floquet engineering. In optical tweezers, typically spaced by a few micrometers, strong interactions can be induced by exciting to high-lying Rydberg levels~\cite{Isenhower2010,Wilk2010}. Contact interactions generally translate to infinite-range interactions in artificial dimensions. In the case of repulsive bosonic atoms in 1D momentum-space lattices, they result, however, in an effectively attractive, finite-ranged interaction, which can be understood via the extra repulsive exchange energy of a symmetrized two-body wave function for distinguishable momentum states~\cite{An2018PRL}. Therefore, interesting many-body physics, such as solitons, are expected in these systems. In the case of synthetic dimensions formed by internal states, it has been proposed to recover short-range interactions required for fractional quantum Hall (FQH) liquids by spatially separating the different internal spin states using a magnetic field gradient~\cite{Chalopin2020}. Interactions also decay with distance in an artificial dimension formed by the levels of a harmonic trap~\cite{Price2017}.

\subsection{Photonic systems}
Another powerful quantum simulation approach is due to the confinement, propagation, and emission of photons via tailored light-matter interactions. It significantly benefits from the advancements in nanofabrication technologies, enabling the construction and precise sculpting of materials at the optical wavelength scale. Furthermore, the availability of various optical spectroscopy techniques—such as time-resolved measurements, Fourier transform imaging, and interferometry—paired with state-of-the-art light sources and photodetectors facilitate the injection, probing, and monitoring of photonic states (both amplitude and phase) in real and momentum space.
%
%
\begin{figure*}
    \centering
    \includegraphics[width=0.95\textwidth]{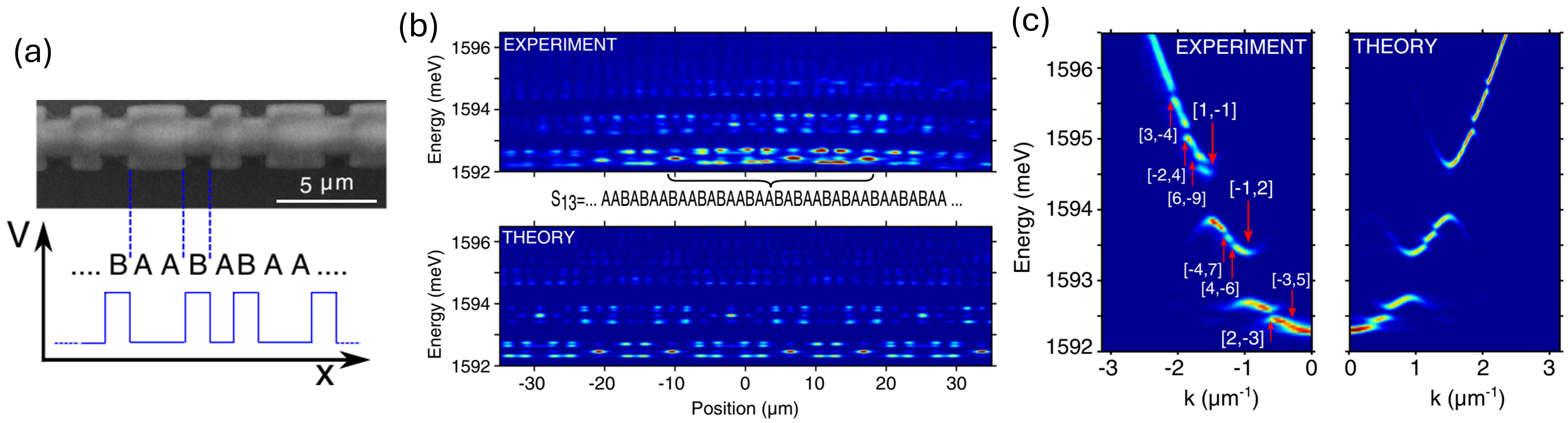}
    \caption{\textbf{Implementation of one-dimensional quasicrystals with photonic systems.} (a) Scanning electron microscopy image of a modulated photonic microwire, showing the shape of the A and B letters of a Fibonacci sequence. (b) Spectrally resolved in real space of emission measured (top) and calculated (bottom) on a microwire. The letter sequence corresponds to a part of the whole $S_{13}$ Fibonacci potential sequence. (c) Spectrally resolved in momentum space of emission measured (left) and calculated (right) on the same wire cavity used in (b). The position of the gaps, dictated by the gap labeling theorem with two integers $[p,q]$, is indicated with red arrows. Adapted from~\textcite{Tanese2014}.}
    \label{fig:Gap_labelling}
\end{figure*}
%
%

Polaritons, quasiparticles arising from the strong coupling regime between confined photons and dipole excitations (electrons oscillating at the metal surface for surface plasmon polaritons and excitons in semiconductors for exciton-polariton), are one of the most successful photonic platforms for quantum simulation. Potential landscapes in 1D are simulated by harnessing superlattices of multilayers along the propagation direction~\cite{ALBUQUERQUE1992383,Poddubny2008,Hendrickson2008}, or by implementing lateral confinement $V(x)$ in photonic microwires through modulating the wire width $w(x)$~\cite{Tanese2014,Nguyen2015,Baboux2017,Goblot2020}. An example of such modulation is shown in Fig.~\ref{fig:Gap_labelling}a. Moreover, simulations of 2D tight-binding Hamiltonians are obtained via lateral coupling between micro pillars~\cite{Jacqmin2014,Sala2015,Klembt2018}. In this scheme, the pillar sizes dictate the on-site energies and the hopping terms given by their inter-distances.  A crucial advantage of the modulated microwires and the coupled micropillar configurations is the out-of-plane access of spectrally resolved photons escaping the microstructures to far-field detectors.  Indeed, a single snapshot of this emission provides tomography of confined polaritonic states in both real and momentum space, as illustrated in Figs.~\ref{fig:Gap_labelling}b,c. Moreover, the pseudospin texture and phase pattern of the polaritonic states can be directly mapped by polarization analysis~\cite{Sala2015} and interferometry measurements~\cite{Sala2015,Jacqmin2014} applied to the out-of-plane emission, respectively. Finally, the matter component of polaritons enables two distinct interaction schemes: (1) self-interaction, facilitating the exploration of many-body physics in the quantum fluid of light~\cite{Nguyen2015}, and (2) interaction with external magnetic fields, breaking time-reversal symmetry for applications in topological photonics~\cite{Klembt2018}.

\begin{figure}
    \centering
    \includegraphics[width=0.45\textwidth]{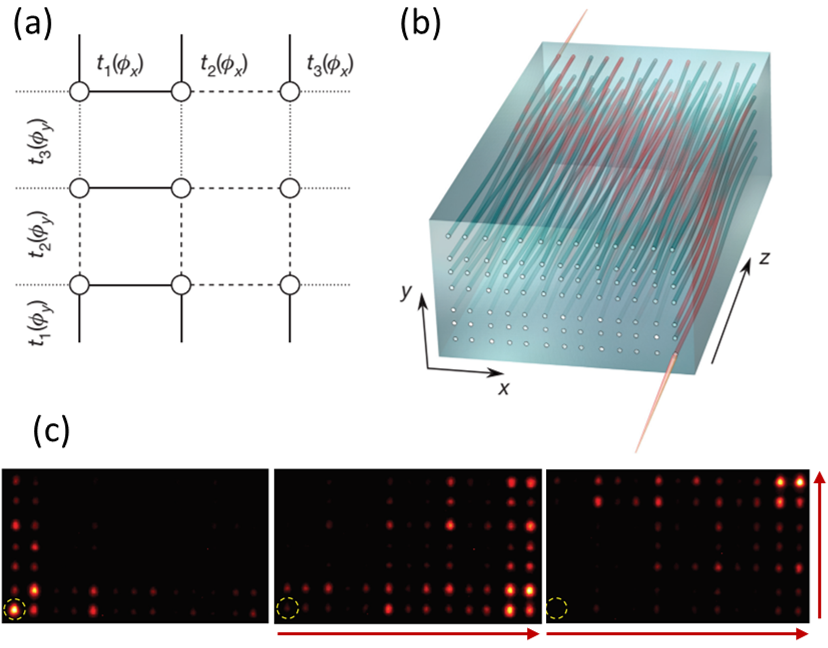}
    \caption{\textbf{Implementation of four-dimensional crystals with photonic systems.} (a) Schematic of a lattice model for 2D topological pump of 4D quantum Hall physics. The hopping terms  are give by $t_n(\phi_x)=t + \lambda\cos\left(2\pi bx + \phi_x\right)$ and $t_n(\phi_y)=t + \lambda\cos\left(2\pi by + \phi_y\right)$. Here the two phases $\phi(x)$ and $\phi(y)$ are two synthetic momenta, playing the role of pump parameters. (b)  Illustration of the 2D array of waveguides with $z$-dependent spacing. Light is injected into the input facet, and is detected on the other side. (c) Images of output facet after 15 mm of propagation with various pump parameters to demonstrate corner-to-corner pumping with from left to right panels: $[\phi_x=0.477\pi,\phi_y=0.477\pi], [\phi_x=2.19\pi,\phi_y=0.477\pi], [\phi_x=2.19\pi,\phi_y=2.19\pi] $. Adapted from~\textcite{Zilberberg2018}. }\label{fig:4D_Photonic}
\end{figure}

Another primary photonic platform for quantum simulations involves arrays of coupled single-mode waveguides. These arrays can be easily fabricated in bulk glass using standard lithography techniques with femtosecond lasers, providing a versatile playground for simulating tight-binding Hamiltonians. This is achieved by engineering the evanescent tunneling of light propagating along a waveguide to its neighbors. Paraxial propagation along the $z$ direction of the waveguide is described by  $i\partial_z \Psi = H \Psi$, equivalent to the Schr\"odinger equation if the third spatial dimension $z$ is equated with time. The onsite and hopping terms of the Hamiltonian are dictated by the refractive index of the waveguides and the spacing between them, respectively. The detection scheme in waveguide array systems is limited to only photographs of near-field intensity in real space. Still, it provides an elegant scenario for monitoring the time evolution of the injected wave packets by capturing the intensity profile at different $z$ positions, thereby studying various transport problems~\cite{Freedman2006,Schwartz2007,Tang2018,Xu2021}. Most importantly, this platform offers a unique scheme to ``dynamically" modulate the Hamiltonian by implementing $z$-dependence in the hopping terms. This paves the way for simulating Floquet physics in topological photonics~\cite{Rechtsman2013,Biesenthal2022} and realizing topological pumping using synthetic momenta~\cite{Kraus2012,Zilberberg2018}, as illustrated in Figs.~\ref{fig:4D_Photonic}a-d.

\subsubsection{Quasicrystals and fractals}
Inspired by pioneering works in solid-state physics on Fibonacci superlattices~\cite{Merlin1985,Todd1986}, early studies on photonic quasicrystals concentrated on Fibonacci superlattices within the polaritonic platform~\cite{ALBUQUERQUE1992383,Poddubny2008,Hendrickson2008}. The first study by~\textcite{ALBUQUERQUE1992383} introduced 2D electron gas (2DEG) sheets at the interfaces of alternating dielectric layers following the Fibonacci sequence and predicted the formation of bulk bands and surface branches of surface plasmon polaritons.  Subsequent research extended this scheme to exciton polaritons in Fibonacci-spaced multiple quantum wells~\cite{Poddubny2008,Hendrickson2008,Werchner:09}. In particular, the experimental results in~\textcite{Hendrickson2008} highlighted the quasicrystal's long-range order, resulting in the formation of a polariton bandgap, and its lack of periodicity, which promotes efficient out-of-plane light emission extraction.

Contemporary research on quasicrystals in polaritonic platforms has predominantly employed the modulated microwire geometry to study topological and critical properties of chains with Fibonacci quasiperiodic potential: The opening of minigaps obeying the gap labeling theorem (see also Sec.~\ref{subsec:topprop})
and the log-periodic oscillations of the integrated density of states have been observed in~\textcite{Tanese2014}. Subsequent work has used localized states at the interface of two Fibonacci polaritonic wires to measure topological invariants associated with the gap labeling~\cite{Baboux2017}. Finally, the critical localization behavior of the Fibonacci chain (see also Sec.~\ref{para:critical_fibonacci}) has been explored in~\textcite{Goblot2020} by continuous interpolation from the AA potential to the Fibonacci, revealing multiple localization/delocalization transitions.

As an alternative to polaritons, waveguide arrays provide another fruitful platform for exploring quasicrystal and fractal physics. Studies of ``stationary" localized states in 1D quasicrystals have experimentally revealed the localization phase transition of the AA model~\cite{Lahini2009} (see also Sec.~\ref{para:loc_AA}). The topological characterization of different 1D quasicrystals via edge states and Chern numbers has been demonstrated in~\textcite{Verbin2013}, as well as the adiabatic pumping of localized edge states within 1D quasicrystals by ``dynamically" sweeping the phase in the AA model, achieved via slowly tuning the spacing between waveguides in~\textcite{Kraus2012}, (see also Sec.~\ref{subsec:topprop}).
Adiabatic pumping provides an analogous behavior to 2D quantum Hall physics within a 1D quasicrystal. Going from 1D to 2D quasicrystals, earl works tackled linear and nonlinear transport and defect dynamics~\cite{Freedman2006}. In this study, quasicrystals were generated by inferential lithography into a birefringent and nonlinear bulk crystal, demonstrating similarities between light propagation and electron tunneling in quasiperiodic potentials.  This study also uncovered soliton formation at high intensities and dislocation dynamics when interactions between crystal sites are permitted. 
Waveguide arrays have also been utilized for the simulation of fractal physics: quantum transport in fractal networks has been explored in~\textcite{Xu2021}, suggesting that anomalous transport is governed solely by the fractal dimension, although theoretical studies show that transport properties can become extremely sensitive to the connectivity at bottlenecks of the fractal~\cite{Rojo-Francas2024}. Finally, by investigating driven fractal lattices within the framework of Floquet physics, fractal photonic topological insulators have been observed experimentally~\cite{Biesenthal2022}. Remarkably, this work demonstrated that topologically protected chiral edge states can exist despite the absence of bulk bands.

\begin{figure}
    \centering
    \includegraphics[width=0.45\textwidth]{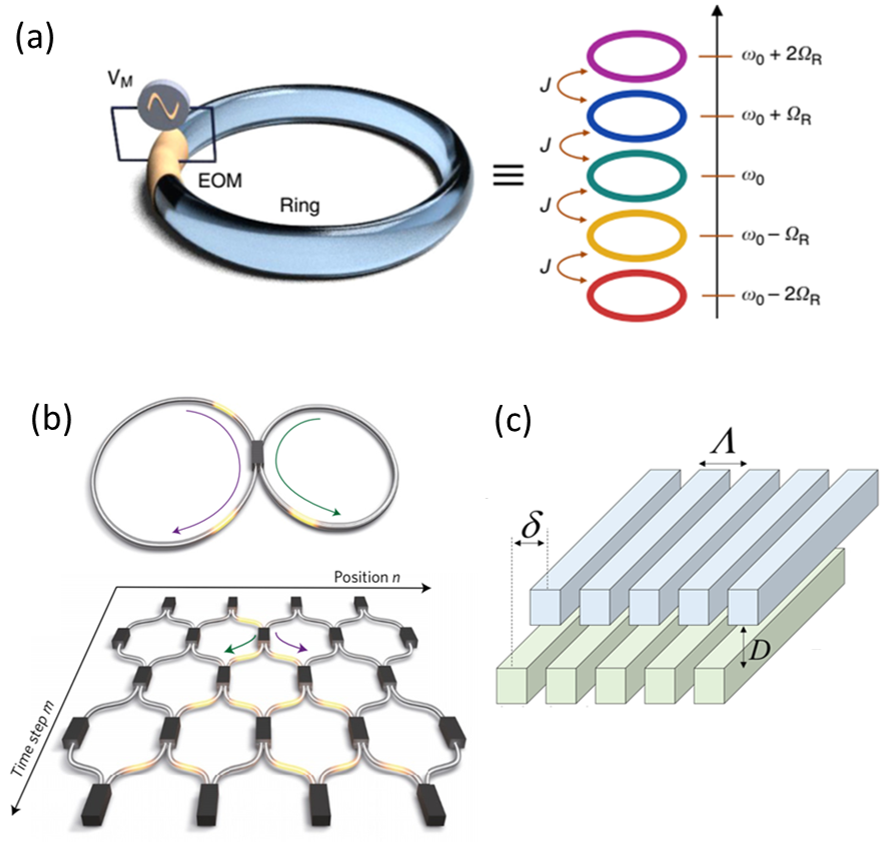}
    \caption{\textbf{Implementation of synthetic dimensions with photonic systems.} (a) Synthetic dimension using frequency lattice. The setup consists of a ring resonator having resonances of equidistant frequency $\Omega_\text{R}$. Periodic modulation with the same period as the round-trip time corresponds to a nearest-neighbor coupling $J$ in the frequency lattice. Adapted from~\textcite{Dutt2019}. (b) Synthetic dimension using time-bin. Light propagating within two coupled fiber loops of different lengths can be mapped onto a quantum walk in a 2D lattice of location and time-step. Adapted from~\textcite{Wimmer2017}. (c) Synthetic momentum. The relative shift $\delta$ between two photonic gratings of the same period $\Lambda$ can be assigned as a synthetic momentum. Adapted from~\textcite{Nguyen2021}.}\label{fig_synthetic_dimension_photon}
\end{figure}

\subsubsection{Lattices with synthetic dimensions}
Photonic systems offer a diverse range of degrees of freedom that can be engineered as synthetic dimensions, enabling the study of higher-dimensional physics using platforms with fewer physical dimensions. A straightforward choice for a synthetic dimension in photonics is the polarization state of light. For example, polarization conversion during light propagation in a photonic lattice can be mapped into an effective coupling between two lattices~\cite{Ehrhardt2021_sciadv}. Other options, based on the nature of photonic modes, include the orbital angular momentum of light~\cite{Cardano2016Natcom,Cardano2017Natcom,Wang2018PRL} or the spatial form of the eigenmodes~\cite{Lustig2022Nature}. In the quantum regime, correlations between photons can be mapped to the synthetic dimension of Fock space. This mechanism can be combined with other degrees of freedom, such as polarization, to achieve more complex mappings~\cite{Ehrhardt2021_sciadv,Ehrhardt2024}.

In addition to using properties of photonic states, another strategy to harness synthetic dimensions in photonics is to employ parameters of the Hamiltonian of a photonic lattice that can be precisely modified. For instance, the physics of the 4D quantum Hall effect can be investigated using 2D quasicrystals, where the two modulation phases $\phi(x)$ and $\phi(y)$ act as synthetic momenta~\cite{Kraus2013,Zilberberg2018}. The lattice model and implementation scheme are depicted in Figs.~\ref{fig:4D_Photonic}a,b, and the experimental results, highlighting corner-to-corner adiabatic pumping, are shown in Fig.~\ref{fig:4D_Photonic}c.

Interestingly, frequency and time-bin, both related to the temporal aspect of space-time, are two of the most commonly used degrees of freedom for synthetic dimensions in topological photonics. Discrete resonances in ring resonators, shown in Fig.~\ref{fig_synthetic_dimension_photon}a, provide a frequency lattice where nearest-neighbor couplings can be engineered on-demand using electro-optical modulation~\cite{Ozawa2016,Yuan2016}. This configuration has been implemented in various material platforms, successfully emulating different facets of topological physics~\cite{Dutt2019,Dutt2020,Wang2021_brading,Balytis2022,Pellerin2024}. Regarding using time-bin as a synthetic dimension, the scheme involves studying light propagation in two loops of optical fiber of different lengths connected by a beam-splitter. This setup makes the system equivalent to a 2D lattice, where one dimension corresponds to the location of the quantum walk and the other to the time step~\cite{Wimmer2015,Wimmer2017,Weidemann2020}, as shown in Fig.~\ref{fig_synthetic_dimension_photon}(b). Additionally, tuning the beamsplitter behavior over time can introduce another synthetic dimension to explore new topological phases, such as the anomalous Floquet metal~\cite{Adiyatullin2023}.

Beyond synthetic coordinates, another essential concept is the synthetic magnetic field (or pseudo-magnetic field). Photons do not interact with magnetic fields directly; therefore, to emulate quantum Hall states in optics, it is necessary to engineer a photonic Hamiltonian that exhibits a synthetic magnetic field. This can be accomplished by deforming a photonic honeycomb lattice, similar to how strained graphene behaves, leading to the experimental observation of Landau levels in coupled micropillar systems~\cite{Jamadi2020} and photonic crystal systems~\cite{Barczyk2024,Barsukova2024}. Combining the ultrastrong nonlinearity of single polaritons with a synthetic magnetic field has ultimately enabled the first realization of bosonic analogs of FQH physics~\cite{Clark2020}. In this setup, single polaritons are formed by the strong coupling between ultra-cold Rydberg atoms and photons in a twisted cavity. This cavity is specifically designed to create a synthetic magnetic field for photons. The observation of angular-momentum-dependent two-photon correlations evidences the creation of Laughlin states.

Synthetic momenta, which differ from synthetic dimensions in spatial space, have recently emerged as a promising concept for exploring higher-dimensional topology. The key idea is to use spatial quantities as synthetic momenta, thereby naturally breaking time-reversal symmetry in the extended momentum space. For instance, in misaligned photonic gratings that share the same period and are evanescently coupled, the relative shifts between the gratings serve as synthetic momenta~\cite{Nguyen2021,Lee2022,Nguyen2023}, as shown in Fig.~\ref{fig_synthetic_dimension_photon}c. Using this concept, chiral edge state~\cite{Nguyen2021} and 2D nodal lines~\cite{Lee2022} were proposed for (1+1)-dimensional systems of bilayer gratings, while Weyl semimetal physics with Fermi arc reconstruction~\cite{Nguyen2023} was proposed for (1+2)-dimensional systems of trilayer gratings. \textcolor{blue}{Recently, synthetic momentum has been employed to explore non-orientable manifolds to circumvent the Nielsen–Ninomiya theorem~\cite{Grossi_e_Fonseca_2024}.} Given the vibrant field of multilayer photonics and various recent realizations~\cite{Tang2023,tang2023onchip}, the experimental implementation of these synthetic momenta systems in the near future is anticipated, potentially addressing 4D or 5D physics.

\subsubsection{Curved lattices}
The propagation of light in a deformed graphene-like lattice of waveguide arrays has been used in~\textcite{Sheng2022} to simulate a gauge field associated with a topological cosmic string. In this experiment, the strain-induced deformation introduces curvature into the effective Dirac equation, which is then mapped onto the lattice Hamiltonian. The degree of curvature experienced by the photons can be tuned by altering the strain strength within the lattice, and it has been demonstrated that the presence of a cosmic string can localize the propagation of wave packets in comparison to flat space.

\subsection{Electronic systems}
%
%
\begin{figure}[!t]
    \includegraphics[width=0.95\columnwidth]{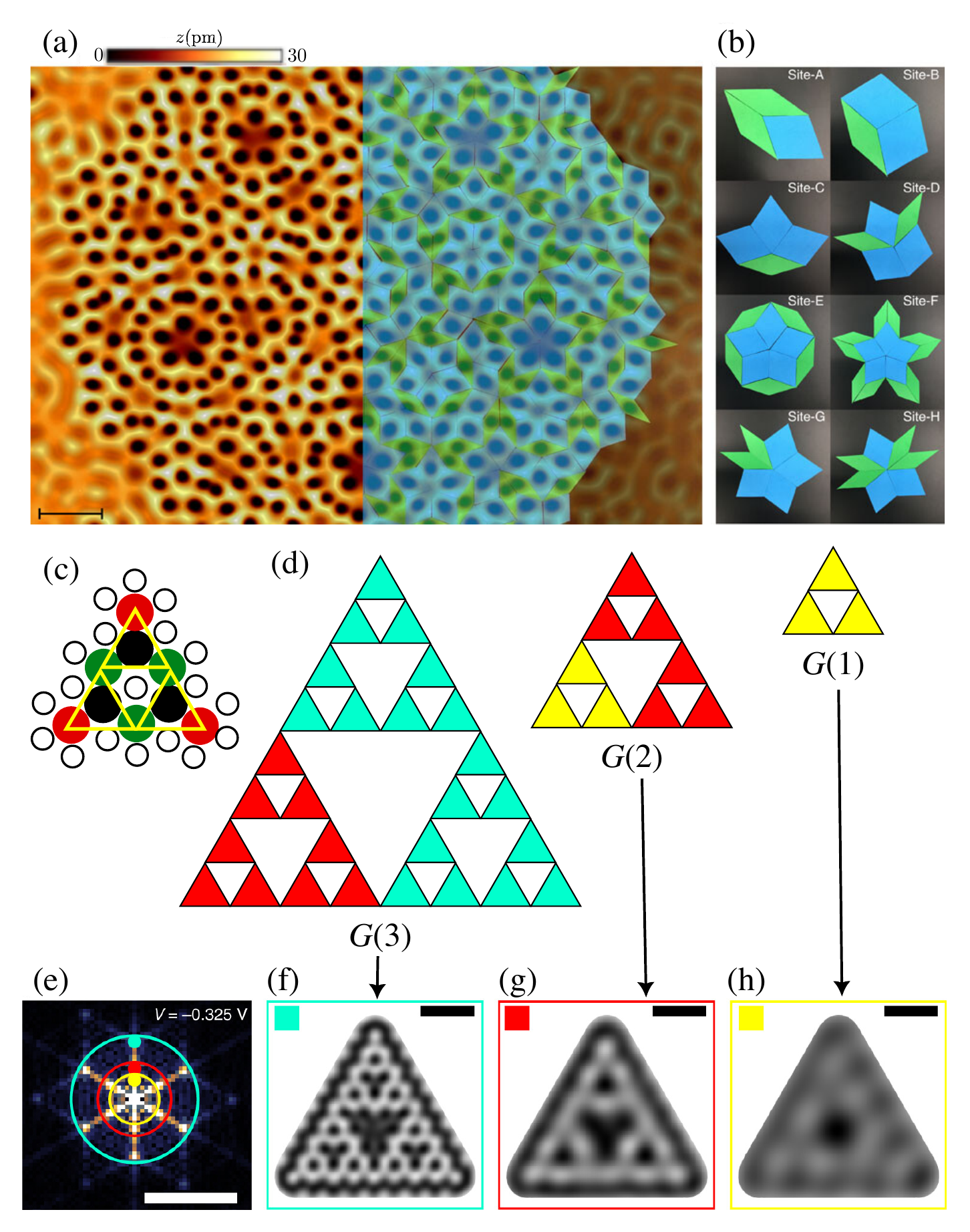}
    \caption{\textbf{Implementation of quasicrystals and fractals by manipulating electrons on surfaces.} (a) STM topograph of assembled quasicrystal composed of 460 CO molecules measured at a bias voltage $V = 10$~mV and setpoint current $I = 1$~nA. The CO molecules are located at the center of each dark spot in the topography. The overlay on the right side is the Penrose tiling composed of rhombi with side length $a_0 = 1.6$~nm and vertices angles $72^\circ/108^\circ$ (blue) and $36^\circ/144^\circ$ (green). Scale bar, $4$~nm. (b) Atlas of the eight types of vertex sites encountered in the Penrose vertex model tiling. Adapted from~\textcite{Collins2017}.  (c) The CO molecules (empty circles) are placed in the G(1) generation configuration. Adapted from~\textcite{Bercioux_2018} (d) The first three generations of the Sierpi{\'n}ski gasket, the colour code shows how the generation $G$($n – 1$) is used in $G$($n$).  (e) Fourier transform of the experimental differential conductance map at $-0.325$~V. The $k$-values outside the circles are excluded from the Fourier-filtered images in (f)–(h). Scale bar: $k = 3$~nm$^{-1}$. (f)–(h) Wavefunction map at $-0.325$~V after Fourier filtering, including merely the $k$-values within the turquoise (f), red (g), and yellow (h) circles indicated in (e). Scale bar: 5~nm. Adapted from~\textcite{Kempkes_2019_b}.
    \label{fig_quasicrystal}}
\end{figure}
%

The last platform we will consider in this section consists of creating artificial lattices in electronic-based systems. We will mainly introduce five different approaches in electronic platforms: (i) electrons on surface, (ii) molecular self-assembly, (iii) twisted layers of two-dimensional materials,
(iv) classical electric circuits, and (v) circuit quantum electrodynamics (QED). In these cases, the goal is to simulate a specific lattice model. 
Some of these platforms are limited to simulate single-particle physics.

In case (i) of electrons on a surface, the electrons of a 2DEG are forced to a specific potential with a specific geometry. Usually, the Shockley surface states of metals are considered for the 2DEG~\cite{Shockley_1939}, \emph{e.g.},  Cu, or Ag on the (111) surface. 
The lattice potential is created by placing atoms or molecules with atomic precision using the tip of a scanning tunneling microscope (STM). The tip of the STM is successively used for measuring the single-particle local density of states (LDOS) $\rho(E)$ at the different $(x,y)$ positions by the differential conductance $dI/dV$ of the system as a function of the applied tip-substrate bias~$V$. This quantity is proportional to the modulus square of the system wave function at a given position and voltage. The major drawback of this technique is the finite lifetime of the surface electronic states. Adatoms of the surface scatter surface electrons into the bulk of the system, leading to the broadening of the electronic resonances. The broadening at 4.6 K is generally smaller in Ag compared to Cu~\cite{Eiguren_2002}. However, the adatoms placed on Cu, especially CO molecules, are more stable and easier to move on Cu than Ag, making Cu the metallic reference system for creating artificial lattice structures~\cite{Freeney_2022}.
This combination has been employed for designing lattices with translational invariance ~\cite{Gomes_2012,Slot_2017,Slot_2019,Kempkes_2019_a,Freeney_2020,Li_2022}. 

The molecular self-assembly approach deposits molecular precursors on a metallic surface; after annealing, the precursors rearrange themselves in structures that can be more complex than the original precursors, either by following a specific pattern intrinsic to the precursor or because of some property of the substrate. Successively, the structure is analyzed either by STM or by an atomic force microscope~\cite{Piquero-Zulaica_2022}.

While the before-mentioned approaches are, at present, limited to single-particle physics, twisted two-dimensional materials provide tunable lattices that showcase strongly correlated physics. A prominent example is twisted bilayer graphene exhibiting unconventional superconductivity at a magic twist angle~\cite{Cao_2018}. Quite generically, twisting 2D lattices leads to a Moiré superlattice determined by the twist angle, yielding the basis for tunable electronic materials that can be used for quantum simulation~\cite{Kennes_2021}. 
One advantage of this platform is that adding more layers or combining different materials (\emph{e.g.} layers of transition metal dichalcogenides) permits the exploration of quantum simulators with diverse properties.

The classical electric circuits approach is based on Kirchhoff's law in the alternating current regime. The general idea is to represent an electrical circuit by a graph in which the nodes and edges correspond to the circuit junctions and connecting elements. The elements of the circuit can be all linear, such as resistors, capacitors, and inductances. The current conservation can be expressed in a compact form as
%
%
\begin{align}\label{Kirchhoff}
    \bm{I}=J(\omega)\bm{V},
\end{align}
%
%
where $\bm{I}$ and $\bm{V}$ are vectors of the input electrical current and voltage at all nodes $a$, here, the Laplacian $J(\omega)=(D-C+W)(\omega)$ is defined via the matrix of adjacency of the conductances $C$, the list of the total node conductance $D$, and the circuit ground $W$. The impedance to the ground of the node $a$, $Z_a(\omega)=V_a/I_a$ can be fully determined by $J$ with eigenvalues $\mathcal{E}_n\propto\omega^{-1}$~\cite{Lee_2018}.

In addition to capacitors and inductances for the circuit QED approach, the circuit will contain non-linear elements such as Josephson junctions. In this case, it can be mapped into an effective many-body Hamiltonian, including the Jaynes-Cummings Hubbard or simpler hopping models~\cite{Schmidt_2013}. Circuits of this type operate in microwave frequency regimes. 

\subsubsection{Quasicrystals and fractals}

\paragraph{Electron on surface.}
A finite portion of the Penrose tiling has been realized using this technique~\textcite{Collins2017}, see Fig.~\ref{fig_quasicrystal}a.   The various rhombi constituting this quasicrystal have been obtained by placing a CO molecule at the center of each of them, see Fig.~\ref{fig_quasicrystal}b. The spectral properties of the system have been analyzed by measuring the LDOS on the various tilings of the quasicrystal. On the different tilings, the LDOS is different, reflecting the distinct nature of the wave function over the quasicrystal.

%
\begin{figure*}
    \includegraphics[width=0.95\textwidth]{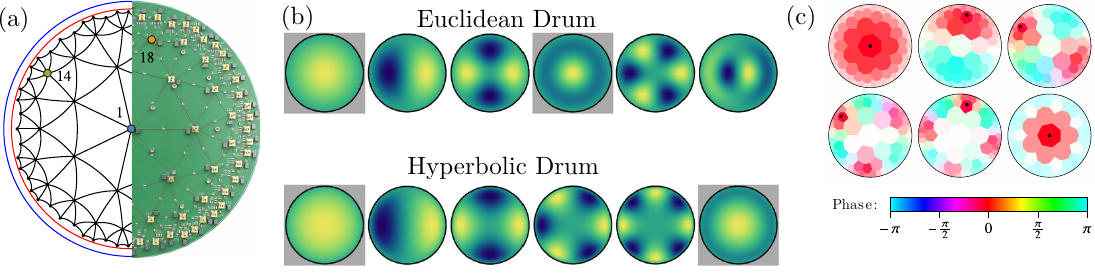}
    \caption{\textbf{Implementation of hyperbolic lattices with classical LC electric circuits.}(a) Schematic of $\{3,7\}$-hyperbolic tessellation (left half) with the unit circle in blue and the circle with radius $r_0 = 0.94$ in red, and photographs of the electric circuit (right half). (b) Comparison of the first six eigenmodes of the Euclidean and hyperbolic drum of radius $r_0 = 0.94$ according to increasing eigenvalues. (c) Measure the voltage profile of the first six eigenmodes (only one mode is shown for each pair of degenerate modes). The saturation encodes the magnitude as a fraction of the voltage (white denotes 0 and full saturation 1) at the input node (black dots), and the color encodes the phase relative to the reference voltage (see legend).  Adapted from~\textcite{Lenggenhager2022}.
    \label{fig_hyperbolic}}
\end{figure*}
%
%

In~\textcite{Kempkes_2019_b}, the electron-on-surface method has been used to implement three generations of the Sierpi{\'n}ski gasket, see Figs.~\ref{fig_quasicrystal}c,d. The experiment directly measured the fractal dimension using the box-counting method. For the Hausdorff dimension of the electronic wave function, a value $\sim$1.58 was obtained at all energies, compatible with the Hausdorff dimension of the Sierpi{\'n}ski triangle, and in contrast to the value of 2 obtained for electronic wave functions in a square lattice~\cite{Kempkes_2019_b}. 
Additionally, the real space analysis of the wave function for different energies displays a standing wave pattern originating from the interference of the electrons scattered by the CO molecules. The Fourier analysis of the wave functions follows the three iterations of the Sierpi{\'n}ski triangle. In other words, the self-similar nature of the fractal in the real space is also present in momentum space, see Figs.~\ref{fig_quasicrystal}e-h.

Some recent research work has shown an alternative method for realizing Sierpi{\'n}ski gasket based on the deposition of Bi on InSb(111)B~\cite{Liu_2021,Ohtsubo_2022,Canyellas_2023}. The interest in this platform is related to the semiconducting nature of the InSb(111)B substrate, leading to a lower broadening of the electronic levels in the simulated lattice system~\cite{Khajetoorians_2019}. The analysis via STM has revealed the formation of various generations of the Sierpi{\'n}ski gasket~\cite{Liu_2021,Ohtsubo_2022}. A recent experiment in~\textcite{Canyellas_2023} has shown the presence of topological modes in these fractals; theoretical calculations indicate that the nature of these topological modes roots down to the intrinsic spin-orbit coupling induced by the fractal confining potential. The same model shows that these topological modes are suppressed by including a competing spin-orbit coupling of Rashba type~\cite{Canyellas_2023,Bercioux_2015}.

\paragraph{Molecular self-assembly.}  Quasicrystals and fractals can also be obtained via molecular self-assembly on a surface. Quasicrystal structures resembling the Penrose tiling have been obtained by depositing C$_{60}$ molecules on a quasicrystal substrate as Al-Cu-Co~\cite{Smerdon_2014}, Al-Cu-Fe~\cite{Fournee_2014} and Bi on Ag-In-Yb~\cite{Hars_2018}.
In the case of the fractals, the first notable success is the work of~\textcite{Newkome2006} where three iterations of the Sierpi{\'n}ski hexagonal gasket were realized, and STM identified the final structure. In~\textcite{Shang2015}, the Sierpi{\'n}ski gasket has been realized, and the STM has been used to estimate the Hausdorff dimension to a value of $1.68\pm0.01$.

\paragraph{Twisted 2D layers. } Stacking and twisting layers of 2D materials is a natural way to obtain large-scale quasicrystalline structures. A quasiperiodic lattice with 12-fold rotational symmetry is obtained from twisting two hexagonal lattice structures by an angle of 30$^{\circ}$ relative to each other~\cite{Stampfli_1986}. This quasicrystal has been realized with twisted graphene bilayers~\cite{Ahn_2018,Yao_2018}, and twisted transition metal dichalcogenides~\cite{Li_2024}. 
It has then been proposed that adding a further twisted layer would permit the combination of the flat-band physics at small twist angles, and the quasiperiodic structure appearing at larger angles can be combined~\cite{Uri_2023}. Such quasicrystal flat bands have then been observed in twisted trilayer graphene~\cite{Hao_2024}, providing a feasible setup for strongly correlated phases of quasicrystals.

\paragraph{Circuit QED.} 
Superconducting qubits can implement Bose-Hubbard models with programmable on-site chemical potential and programmable hopping terms. This has been used in~\textcite{Roushan2017, Li_2023} to build short chains (up to 10 sites) with quasiperiodic potential and/or tunneling, implementing 1D quasicrystals of bosons.

\subsubsection {Lattices with curvature} 
Lattices with negative curvature have been simulated within circuit QED and classical electric circuits.

\paragraph{Circuit QED.} 
In~\textcite{Kollar2019}, a hyperbolic kagome lattice with Schl\"afli numbers $\{7,3\}$ has been realized in coplanar waveguide resonators coupled to each other via superconducting qubit.
Here, the authors have shown that the theoretical transmission spectrum of the circuit is compatible with a $\{7,3\}$-hyperbolic lattice~\cite{Boettcher_2020}, with $\mathcal{D}=5$. This lattice is one of the possible hyperbolic analogs of the kagome lattice and also presents a flat band. For $p$ odd, it presents an energy gap between the flat band and the remaining bands absent in the Euclidean case. 

\paragraph{Classical electric circuits.} In~\textcite{Lenggenhager2022}, the normal modes of Euclidean and hyperbolic lattices have been studied. Specifically, the authors have implemented a $\{3,7\}$ lattice with $\mathcal{D}=5$ and a $\{3,6\}$ lattice with $\mathcal{D}=4$. They have identified a clear difference between the Euclidean and the hyperbolic drum eigenmodes of the two lattices by evaluating the difference in the propagation of waves inside the lattice, see Fig.~\ref{fig_hyperbolic}. 
A similar experimental implementation in~\textcite{Zhang_2022} has proposed two different topological implementations of a $\{6,4\}$ hyperbolic lattice. In the first one, the authors implemented a hyperbolic version of the Haldane model, showing the existence of chiral edge modes. In the second one, the authors implemented a dimerization of the hopping terms to study higher-order modes; within this approach, they have shown a fractal-like behavior of midgap higher-order zero modes~\cite{Pai2019}. In a more recent implementation, it has been possible to implement an $\{8,3\}$ hyperbolic circuit with periodic boundary conditions and tunable complex phases, as well as hyperbolic graphene lattices showing the Dirac cones~\cite{Chen_2023}.

%
%
\begin{figure}[!h]
    \centering
    \includegraphics[width=\columnwidth]{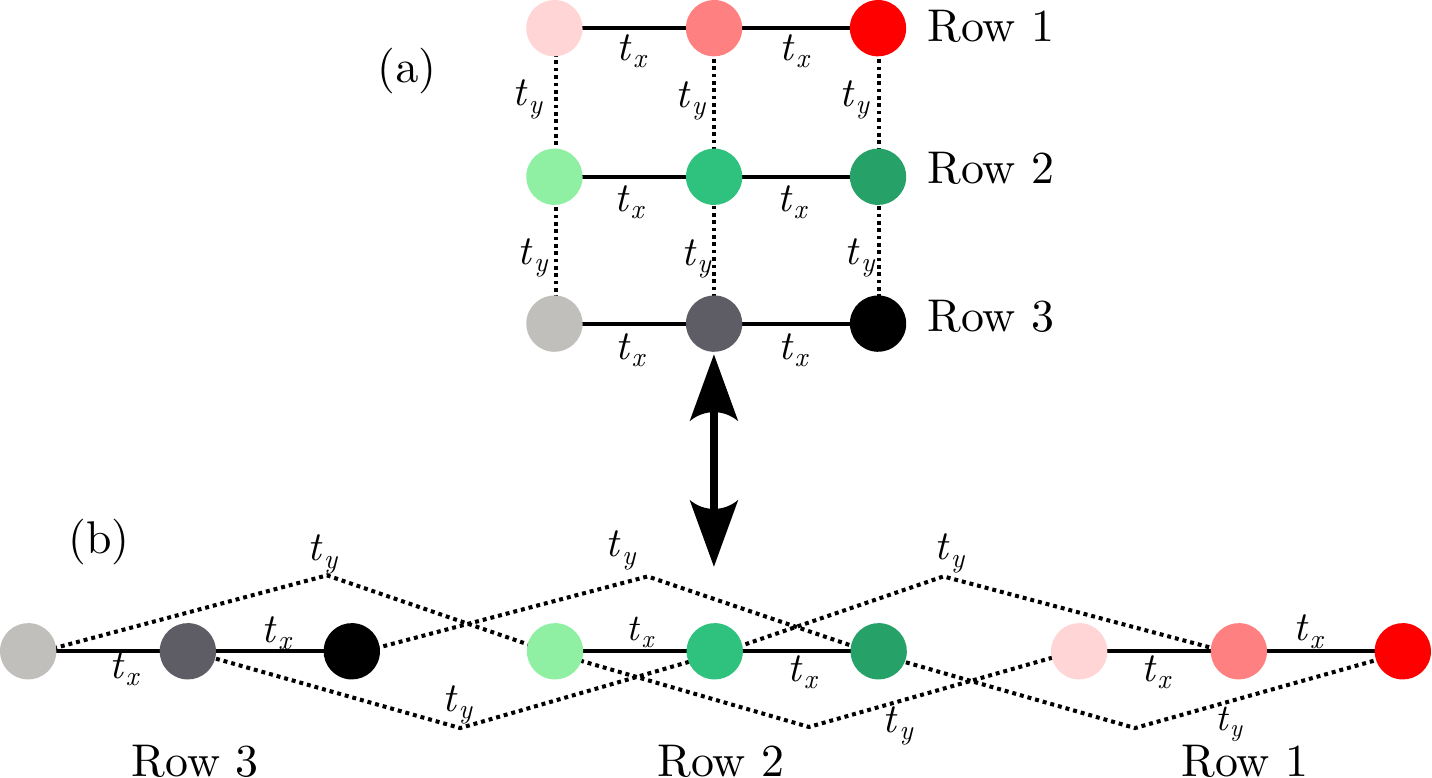}
    \caption{\textbf{Embedding of lattices into lower dimensions.} (a) The sites of a 2D lattice with hopping terms $t_x$ and $t_y$ are rearranged into a 1D chain (b), where the hopping terms $t_y$ are now of long-range type while keeping the connectivity of the original 2D lattice. This embedding strategy was used to realize an effective 4D lattice in a 3D stack of printed circuit boards in~\textcite{Zhao_2018}.}
    \label{Fig_2D_1D}
\end{figure}
%
%
\subsubsection{Higher dimensional physics}
Classical electric circuits are also a versatile platform for implementing higher-dimensional physics. The two-dimensional array of capacitors and inductors can be easily extended to realize periodic 3D and 4D circuits on a breadboard~\cite{Zhao_2018}. The idea is to decompose an $N$D structure into a set of connected $(N-1)$D circuit boards; see Fig.~\ref{Fig_2D_1D}.

In~\textcite{Wang2020}, a 4D lattice has been implemented to realize a topological insulator. Here, the 4D lattice system has been implemented using capacitive and inductive connections fixed in a specific way to implement nonlocal connections in 3D space. Spinless particles with preserved time-reversal symmetry are simulated. Thus, the system is in class AI according to the 10-fold classification of topological metter~\cite{Ryu2010}. Within this classification, a system in class AI presents a topological phase in 4D.
Impedance measurements reveal the signature of corner modes associated with this topological phase. 
In~\textcite{Yu_2020}, a similar setup is theoretically investigated, and in addition to corner modes, the authors also predict topological signatures as 3D Weyl states.

\subsection{Summary}
As demonstrated by the above sections, for each of the exotic geometries considered in this colloquium, various possible implementations exist that use atoms, photons, or electrons. Despite the versatility of each platform, they have strengths and weaknesses, which we outline in Tab.~\ref{tab:sw}. Thereby, one should bear in mind that, at least to some extent, these categories are rather subjective, and a nuisance in one context can be an opportunity in another one.

\begin{table*}
\centering
\includegraphics[width=0.98\textwidth]{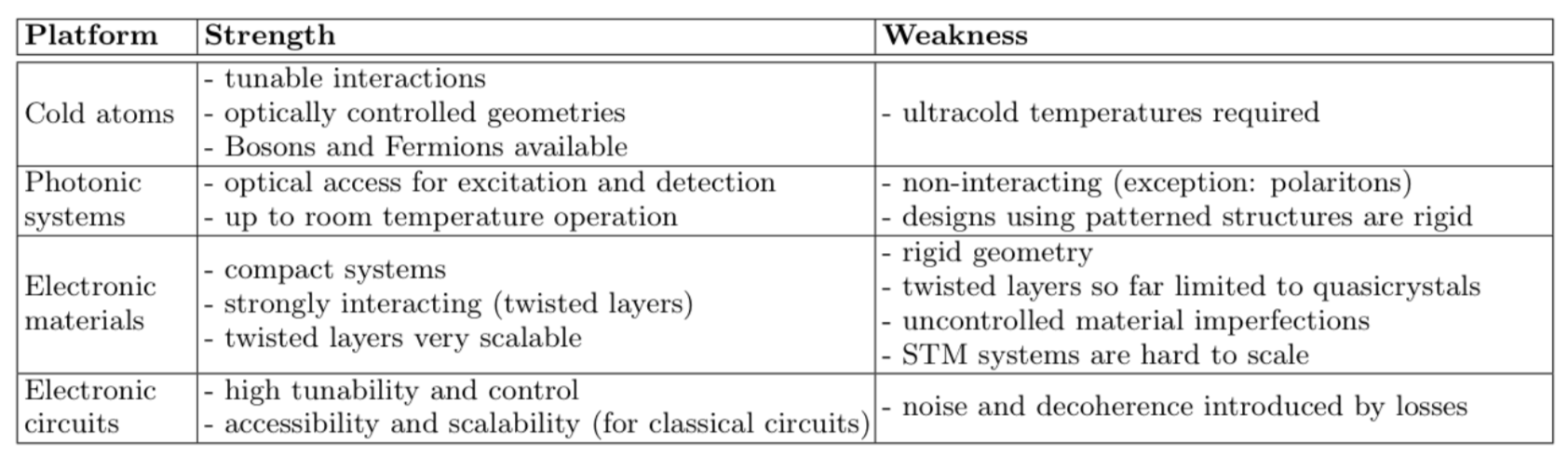}
\caption{Strengths and weaknesses of various platforms.}
\label{tab:sw}
\end{table*}

\section{Applications - novel phenomena in exotic geometries  \label{Part2}}

This colloquium section shall provide an overview of the fascinating topics that can be addressed by implementing the different types of exotic lattices discussed in the previous section.
These phenomena interest researchers from various fields, from condensed matter physics to cosmology.

\subsection{Localization phenomena \label{Sec:loc}}
\subsubsection{ Single-particle localization}
\textcite{Anderson1958} demonstrated that single electrons in uncorrelated static disordered materials exhibit vanishing diffusion at long times, with an exponentially localized wavefunction. The key mechanism in 1D and 2D systems is destructive wave interference everywhere except at the initial point, leading to inevitable localization even in the presence of infinitesimal disorder. However, in 3D, complete backscattering is less likely, requiring a stronger disorder strength for localization to occur, and giving rise to a metal-insulator transition at a critical disorder strength. The transition is characterized by a power-law divergence of the localization length near the critical energy, that is, the mobility edge.
In recent times, the phenomenon of localization has also been pursued in the context of quasiperiodic potentials, revealing a rich array of phenomena that challenge our conventional understanding. 
Compared to disordered systems, quasiperiodic systems display a more diverse localization behavior, including multifractal states and mobility edges even in one dimension. This diversity arises from the unique interplay between disorder and the inherent periodicity of quasiperiodic structures, providing an alternative perspective on the fundamental mechanisms of localization. 

\paragraph{Measures of localization.} Let us first introduce some tools to quantify localization properties. A simple and frequently used quantity is the inverse participation ratio $R_\alpha$ of a state $|\alpha\rangle$
%
%
\begin{align}
R_{\alpha} &= \sum_{j=1}^N |u_{j,\alpha}|^4,
\end{align}
%
%
with $u_{j,\alpha}=\langle j| \alpha\rangle $ being the amplitude of the state $|\alpha\rangle$ at site $j$, and $N$ being the total number of sites in the system.  The mean inverse participation ratio $R$ is obtained by averaging over all eigenstates, $R = \frac{1}{N} \sum_\alpha R_\alpha$. For a perfectly localized system, with all states pinned to a single specific site, $R$ becomes 1, whereas for perfectly extended (delocalized) states $R=1/N$. Hence, the scaling of $R$ with $N$ is a useful indicator for differentiating localized/extended states. Another classification tool is the $q$-fractal dimension $D_q$~\cite{Halsey1986,Evers2008} of the eigenstates. For a specific state, this quantity is defined by the following relation $\sum_{j=1}^N |u_{j,\alpha}|^{2q} = N^{-D_q(q-1)}$. For $q=2$, the $q$-fractal dimension is proportional to the logarithm of the inverse participation ratio of a state. Averaging over all eigenstates, the mean fractal dimension $\bar D_2$ approaches the value 1 (0) for extended (localized) systems, whereas intermediate values suggest multifractal behavior. The information carried by the $q$-fractal dimensions $D_q$ can also be expressed through the multifractal spectrum $f(\alpha_q) \equiv q(\alpha_q) - \tau_q$, where $\tau_q \equiv D_q (q-1)$ and $\alpha_q \equiv \frac{\rm d}{{\rm d}q}\tau_q$~\cite{Halsey1986}.  

\paragraph{Localization properties of the AA model. \label{para:loc_AA}}
The AA model (see Eqs.~\eqref{AAH} and \eqref{AA}) exhibits all three phases (extended, localized, and critical)~\cite{Tang1986,Hiramoto1989}. For disorder strength $|\lambda|>2|t|$ ($|\lambda|<2|t|$), all eigenstates are localized (extended). The eigenfunctions are critical or multifractal at $|\lambda|=2|t|$.

Without explicit calculations, some of the localization behavior of the AA model can be understood from a duality relation: Consider the Fourier-like transformation $a_n = \sum_k \tilde a_k \exp(i 2\pi \alpha n k)$ which transforms the real-space localized operator $a_n$ into a fully delocalized one, and vice versa. Under this transformation, the AA Hamiltonian (with $\phi=0$) becomes
%
%
\begin{align}
H_{\rm dual}= \sum_k \left[ -2t \cos(2\pi\alpha k)  \tilde a_k^\dagger \tilde a_k + \frac{\lambda}{2} \tilde a_{k+1}^\dagger \tilde a_{k} + {\rm h.c.} \right]. 
\end{align}
%
%
For $|t|/|\lambda|=1/2$, the model is self-dual, that is, the Hamiltonian does not change its form under the duality transformation. At this point, all eigenfunctions must have the same distribution in both real and momentum space.

The localization properties of the AA model have been probed using quasiperiodic photonic waveguide lattices~\cite{Lahini2009} and ultra-cold atoms~\cite{Roati2008}. The localization transition was identified by analyzing transport, spatial and momentum distributions, and confirmation of scaling behavior for critical disorder strength. These experiments unequivocally demonstrated the transition from extended to localized states as the strength of the quasiperiodic potential increases.

\paragraph{Generalized AA models with mobility edges.}
Like 1D disordered systems, the AA model has no mobility edge, and all states in the spectrum behave similarly, i.e., they are either localized or delocalized, irrespective of their energy. 
However, a 1D quasiperiodic system generally may have mobility edges if the self-duality condition mentioned earlier becomes energy-dependent. In this case, all eigenstates up to a certain energy are localized, whereas states above this ``mobility edge" are extended. Different modifications of the AA potential are known to produce a mobility edge~\cite{Soukoulis1982,DasSarma1988,Hiramoto1989,Ganeshan2015}, for instance: 
%
%
\begin{equation}\label{VgAA}
V_n= \lambda \frac{\cos(2\pi\alpha n +\phi)}{1-\beta \cos(2\pi\alpha n +\phi)}. 
\end{equation}
%
%
It contains the AA potential of Eq.~\eqref{AA} as a limiting case ($\beta=0$) but yields a model with a mobility edge away from this limit. Generalized dual transformations can also be applied to generalized AA (GAA) models with mobility edge~\cite{Ganeshan2015}. Other ways to obtain a mobility edge in an AA-like model include hopping beyond nearest-neighbors~\cite{Biddle2011,Deng2019}. An additional richness to the localization properties of 1D quasicrystal can be obtained by staggered hoppings, as in the Su-Schrieffer-Heeger model~\cite{Su1979}. Increasing the quasiperiodic modulation can immediately delocalize already localized states, giving rise to re-entrant localization transitions~\cite{Roy2021}. 

Following~\textcite{Hinrichs2007,Li2017}, a 1D quasicrystal with an energy-dependent mobility edge has been implemented with cold atoms in an optical lattice, see~\textcite{Lueschen2018}. An alternative atomic realization of the generalized AA model has been presented in~\textcite{An2021} using a momentum-space lattice. It has also been shown that a balanced bichromatic optical lattice, where the amplitude of both frequency components are comparable, leads to a model with a mobility edge, beyond the standard AA model.

\paragraph{AA model with pairing term.} 
The presence of pairing terms modifies the localization behavior of the AA model. In this case, the quasiperiodic modulation can also affect the pairing, as studied in~\textcite{Yahyavi2019}. It has been shown that topological superconductivity can co-exist with critical localization properties. While in the absence of pairing the AA model is multifractal only at the critical point of the localization transition, the pairing gives rise to a broader critical regime~\cite{Fraxanet2022}. The phase diagram becomes even richer in the presence of long-range hopping and/or long-range pairing, where the spectrum exhibits a crossover between localized, multifractal, and ergodic regions~\cite{Fraxanet2022}. 
A mixed spectrum has also been seen in the case of the 2D AA-like model, exhibiting extended states on both sides of the self-dual point~\cite{Szabo2020}. 

\paragraph{Critical behavior of the Fibonacci chain. \label{para:critical_fibonacci}}

In contrast to the AA model, the Fibonacci chain (see Sec.~\ref{Intro:quasicrystals}) does not exhibit a localization transition and remains critical~\cite{Ostlund1983,Kohmoto1983} for any value of the modulation strength as dictated by the Luck criterion of bounded fluctuations in the pattern~\cite{Luck1993}. While the fluctuations can become arbitrarily large for systems with uncorrelated noise, fluctuations are bounded for many quasicrystals, including the Fibonacci quasicrystal. Systems with bounded and unbounded fluctuations have been shown to belong to different universality classes. 
Using transfer matrix method~\cite{Benza1990}, scaling analysis~\cite{Luck1993}, real-space renormalization group~\cite{Igloi1997,Agrawal2020}, the critical behavior of quantum Ising chains with Fibonacci modulation has been studied, establishing conditions for the occurrence of magnetic phase transitions, and for the quasiperiodic system to be in a different universality class than the uniform one~\cite{Luck1993,Agrawal2020}.  
Also the critical behavior of the Heisenberg model, $H_{XXX}= \sum_{i} J_{i} {\bf S}_i \cdot {\bf S}_{i+1}$ with quasiperiodic couplings $J_{i}$ has been studied~\cite{Hida1999,Hida2004,Igloi2007,Agrawal2020}, as well as the XXZ model $H_{XXZ}=\sum_{i} J_i (S_i^x S_{i+1}^x + S_i^y S_{i+1}^y + \Delta S_i^z S_{i+1}^z)$~\cite{Vieira2005}. 
 It has been shown that a Fibonacci modulation of the Heisenberg coupling alters its critical behavior, in contrast to the XY model or the quantum Ising model, where a quasiperiodic modulation of the coupling is marginal or irrelevant as the central charge remains unaffected.
 However, we come to these later since these spin models map to interacting fermionic models.

\paragraph{Localization in 2D quasicrystals.} Non-interacting systems in lower-dimensional landscapes, defined by static, random potential energy and short-ranged tunneling, exhibit Anderson localization. Transitioning to higher dimensions facilitates delocalization, driven by increased energy density. For quasicrystals, with the increasing sophistication of experimental techniques, research into localization has extended to two-dimensional (2D) quasicrystals, revealing a multifaceted landscape of localization behavior. Quasicrystalline order usually arises from projecting a higher-dimensional periodic lattice in an incommensurate manner. Penrose introduced a method for constructing quasicrystals using a set of tiles and matching rules, resulting in the well-known fivefold symmetric Penrose tiling and the eightfold symmetric octagonal tiling~\cite{Penrose1974}. These tilings, with disallowed rotational symmetries, exhibit self-similarity in both real and reciprocal space. Self-similarity upon scaling indicates nontrivial structure at arbitrarily large scales, reflected in quasicrystal diffraction patterns with sharp peaks at very small momenta. In the studies~\textcite{Viebahn2019,Sbroscia2020}, a BEC of Potassium atoms is utilized to probe a 2D quasicrystalline optical lattice through stimulated two-photon Kapitza-Dirac scattering. The quasicrystalline potential is created by combining four mutually incoherent 1D optical lattices, forming a global eightfold rotational symmetry. Unlike periodic lattices, the combination of reciprocal lattice vectors results in new, smaller momentum scales, creating a self-similar fractal structure (see Fig.~\ref{fig:implementation-quasicrystals-fractal}b). This process is akin to the incommensurate projection of a 4D simple cubic lattice to the 2D plane, similar to the construction of the octagonal tiling. Adjusting the number of lattice beams allows control over the dimensionality of the parent lattice. 
It has been predicted that the spectrum consists of a localized low-energy part, an extended high-energy part, and an intermediate regime, in which localized and critical states alternate~\cite{Zhu2024}.
Finally, adiabatically loading BECs into quasicrystal lattices allows studying Bose glasses in this geometries~\cite {Yu2023}. 
The system thermalizes at higher energy density and weaker disorder, exhibiting a conducting phase.

\paragraph{Localization in fractal lattices.}
Also, fractal structures, such as the Sierpiński gasket, are known to host localized eigenstates~\cite{Domany_1983}, in addition to an infinite number of extended states~\cite{Wang_1995}. This can give rise to complex transport behavior, featuring an increased  return probability in quantum walks as compared to classical random walks~\cite{Darazs_2014}.
Recently, quantum transport in fractal geometries has been explored in waveguide experiments~\cite{Xu2021}, and super-diffusive quantum transport has been observed in both Sierpiński gasket and carpet. However, in the experimental setup connectivity between all nearest neighbors has avoided a bottleneck in the Sierpiński gasket, which occurs at sites where two fractal generations are glued together. As pointed out by Ref.~\textcite{Rojo-Francas2024}, this bottleneck significantly affects the energy spectrum of the system, and slows down quantum transport into the sub-diffusive regime. Theoretical studies of fractal lattices with random connectivity have shown that eigenfunction localization depends on the spectral dimension rather than the Hausdorff dimension~\cite{Kosior2017}.

\paragraph{Localization in higher-dimensional or curved spaces.} While quasicrystalline or fractal geometries may give rise to localization in the absence of disorder, quantum particles in Euclidean 1D or 2D lattices are localized by an infinitesimal amount of disorder. A different behavior has been reported for hyperbolic lattices, where finite disorder is required for localization~\cite{Chen_2024}, similar to the localization behavior in higher dimensions $d \geq 3$, see~\cite{Tarquini_2017}. Quantum funneling, a localization effect caused by the singularity in a negatively curved space, has been described by~\textcite{Zhang_2021}. In the future, quantum simulators of exotic geometries may be used to probe localization behavior in higher-dimensional or curved spaces.

\subsubsection{Many-body localization}
Quantum simulators of quasicrystals have also yielded groundbreaking discoveries in the many-body context, showcasing their ability to test theoretical predictions experimentally. Traditionally, a disordered potential localizes single-particle quantum eigenstates, rendering the system an insulator with zero conductivity. In~\textcite{Basko_2006}, it was proposed that interacting many-body systems could similarly undergo a MBL transition. This transition involves highly excited states, occurring at low energy density and strong disorder, where the system remains a perfect insulator. The system thermalizes at higher energy density and weaker disorder, exhibiting a conducting phase. MBL violates the eigenstate thermalization hypothesis (ETH), which posits that each eigenstate at finite energy density has the same expectation
values of local physical observables as the ones provided by local thermal states. This implies that thermalization occurs at the level of individual eigenstates in an isolated quantum system, where no memory of the initial state survives. The eigenstates of the system in the MBL phase do not obey the ETH, and some memory of the initial conditions is retained
in local observables for arbitrarily long times.

\begin{figure*}
    \centering
    \includegraphics[width=0.95\textwidth]{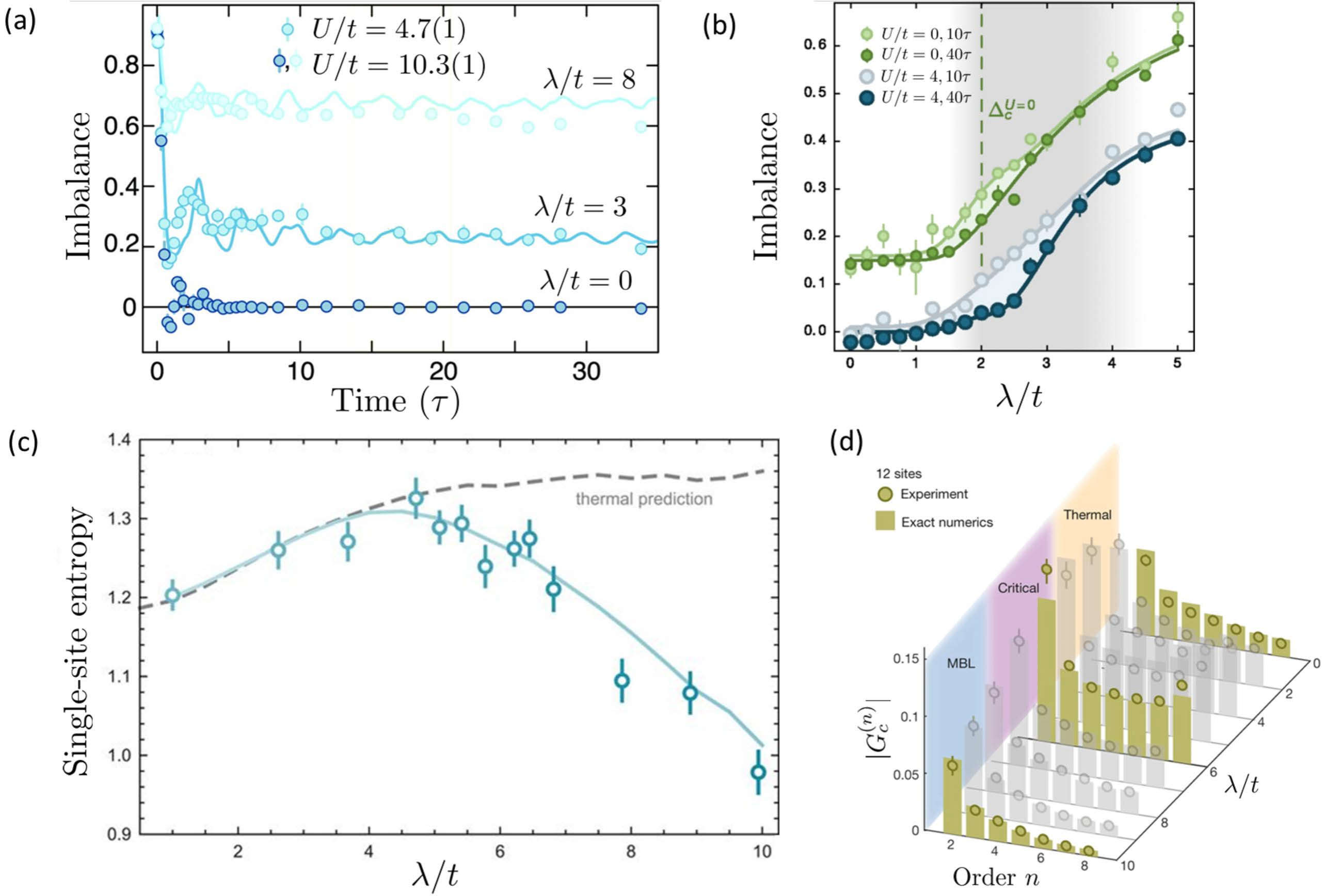}
    \caption{\textbf{Probing many-body localization.} (a) Time-evolution of the population imbalance between even/odd sites for different strengths of the quasiperiodic potential $\lambda/t$, in good agreement with theoretical simulations using DMRG (solid lines). Initializing the experiment in a strongly imbalanced state, the imbalance initially drops and saturates at a value that depends on $\lambda/t$. (b) The value of the imbalance after 10 (40) tunneling times $\tau=\hbar/t$ is plotted as a function of the quasiperiodic potential strength $\lambda/t$ for the non-interacting system (green) and an interacting system (blue). For clarity, the non-interacting data in (b) is vertically offset by 0.15.
    Due to the harmonic trap, the onset of localization is shifted to values slightly below $\lambda/t=2$ in the non-interacting systems, whereas interactions shift the localization towards larger values of $\lambda/t$. Note that in the shaded region, the interacting system has not yet saturated after a time 10$\tau$, suggesting that it is a regime of critically slow relaxation. The data in (a,b) is taken from the Munich experiments, adapted from~\textcite{Schreiber2015,Lueschen2017}. (c) A reduced single-site von Neumann entropy indicates the MBL phase in the Harvard experiment, adapted from~\textcite{Lukin2019}. The small system size gives rise to a broad quantum-critical regime with strong multi-site correlations captured by n-point connected correlations $\vert G_c^{(n)}\vert$, as shown in (d), adapted from~\textcite{Rispoli2019}.}
    \label{fig:MBL_exp}
\end{figure*}

\paragraph{Theoretical investigations.}
Various theoretical studies have considered localization on the many-body level in quasiperiodic potentials. In fact, the quasiperiodic scenario with AA modulation and the purely random scenario have explicitly been compared in~\textcite{Khemani2017}, and it has been argued that the quasiperiodic potential produces a more stable MBL phase than the quenched disorder. Moreover, it has been claimed that the two systems belong to different universality classes. A huge number of theoretical works addressed the MBL phase of interacting particles in the AA potential~\cite{Iyer2013,Khemani2017,Bera2017,BarLev2017,Setiawan2017,Lee2017,Weidinger2018,Doggen2019,Xu2019,Yao2020}. There is full agreement that, in the case of an AA modulation, the localization transition in interacting systems occurs at larger values of the quasicrystal potential than in the free system. Most of the literature results locate the transition within $3\leq \lambda/t \leq 5$. However, no narrow value exists. One reason for this is that all studies are finite-size studies, which leads to some intermediate regime, cf.~\textcite{Rispoli2019} and Fig.~\ref{fig:MBL_exp}(d). However, in this context, we also note that~\textcite{Doggen2019} claims that finite-size effects are rather weak for MBL systems in quasiperiodic potentials, which is in contrast to the purely random disorder.
On the other hand,~\textcite{Xu2019} studies the scrambling of a localized perturbation and finds that it is suggestive of an intermediate phase.
Another reason why the transition value is not sharply determined is that different figures of merit can be used to determine the localization transition. For instance,~\textcite{Lee2017} reports $\lambda_c=3.7t$ by considering the finite-size scaling of the entanglement entropy, but $\lambda_c=5t$, from the quench dynamics, when the system is initially prepared in a very imbalanced state. 
There have also been theoretical studies of localization behavior in interacting Fibonacci chains~\cite{Mace2019,Varma2019,Chiaracane2021}, as well as in the interpolating AA-Fibonacci potential~\cite{Strkalj2021}. While a non-interacting Fibonacci system remains critical for any potential strength, that is, it does not localize,~\textcite{Mace2019,Chiaracane2021} claim evidence of a crossover to MBL for Fibonacci chains with sufficiently strong interactions. On the other hand, the absence of a MBL phase was reported for a weakly interacting Fibonacci system~\cite{Varma2019}.
Recent theoretical work has also studied MBL on fractal lattices in the presence of random disorder~\cite{Manna2024}. These systems provide an excellent opportunity for exploring the still unknown fate of MBL beyond 1D systems.

In general, the theoretical description of the MBL transition is extremely challenging. In contrast to the metal-insulator transitions, which occur in the ground state or the low-energy states of a system, the MBL transition affects the whole spectrum or at least an extended spectral range, which might be far away from the ground state. For this reason, much of the machinery to study many-body phases of matter (such as Monte Carlo techniques or tensor network methods) is not well suited for studying MBL. As a consequence, numerical explorations have mostly relied on exact diagonalization studies of systems with 20-30 spins, and even these relatively moderately sized studies strongly profit from sophisticated techniques~\cite{Pietracaprina2018,Sierant2020,Sierant_2024}, exploiting the sparseness of the matrix. Larger system sizes have been treated with a self-consistent Hartree-Fock description~\cite{Weidinger2018} or time-dependent density matrix renormalization group (DMRG) studies~\cite{BarLev2017,Xu2019}.

Different fingerprints of the MBL phase are known: (i) The half-chain von Neumann entanglement entropy $S$ is strongly suppressed in the MBL phase~\cite{Lee2017,BarLev2017,Mace2019}. The transition point can be estimated by considering the value at which the entanglement entropy per site $S/L$ is independent of the system size $L$. (ii) The level spacing ratio $r_n={\rm min}(\frac{g_n}{g_{n+1}}, \frac{g_{n+1}}{g_n})$, where $g_n=E_n-E_{n-1}$ is the spacing between adjacent energy levels, differentiates between localized and ergodic phase. It is well established by random matrix theory that the level spacing distribution characterizes the statistical ensemble to which a matrix belongs~\cite{Haake2001}. Ergodic systems typically belong to the Wigner-Dyson ensemble, characterized by level repulsion and a relatively large value $r\sim 0.53$.  On the other hand, integrable/localized systems are characterized by the Poisson statistics, $r\sim 0.39$, see~\textcite{Pal2010}. (iii) Participation ratio and fractal dimension of the eigenstates, as introduced in the Section on non-interacting systems, can also be used for a many-body system if, instead of the full many-body eigenstates, the eigenstates of the one-body density matrix are considered~\cite{Bera2017,Mace2019}. (iv) Dynamics probes have been used in many experiments: If a system is localized, it should keep memory of its initial state during a quench. Often, systems are prepared in a state with alternating empty and occupied sites, and the evolution of this even/odd population imbalance is tracked. In contrast to an ergodic system, the MBL system remains imbalanced on long time scale~\cite{Schreiber2015}.

\paragraph{Experimental studies.}

Given the computational difficulties, quantum simulation has become an extremely valuable tool for studying MBL. At the same time, due to the instability of the MBL phase against a thermal bath, the study of the MBL phase is also motivated by the unique opportunity of isolating quantum systems from the environment offered by quantum simulators. The primary systems where the MBL transition is experimentally studied are cold atoms in optical lattices with an AA modulation~\cite{Schreiber2015,Lueschen2017,Kohlert2019,Lukin2019,Rispoli2019}. Experiments in Munich~\cite{Schreiber2015,Lueschen2017,Kohlert2019} have concentrated on rather larger systems, arrays of 1D tubes with approximately 150-200 sites per tube, filled with spin-1/2 fermions. 

An important figure of merit in these experiments is the population imbalance between even and odd sites. Initially, the system was prepared for a large imbalance, which was then washed out through dynamic evolution. While in the absence of a quasiperiodic potential, the imbalance fully disappears on short time scales, a sufficiently strong quasiperiodic potential leads to saturation of the imbalance at a finite value, see Fig.~\ref{fig:MBL_exp}a. By plotting the saturation value versus the potential strength $\lambda/t$, see Fig.~\ref{fig:MBL_exp}b, the onset of localization is found to be shifted towards larger values of $\lambda/t$ by the interactions. In the non-interacting case, the transition seems to occur slightly below the single-body critical value $\lambda/t=2$, which can be accounted for by the harmonic trapping potential compared with exact numerical simulations. In the vicinity of the localization transition, especially in the interacting system, the saturation occurs only on much longer time scales than deeply in the ergodic or the MBL regime.  \textcite{Kohlert2019} also studied MBL in a GAA model. In contrast to the AA model, the non-interacting GAA model exhibits an intermediate phase in which both localized and delocalized eigenstates co-exist. While the experiment confirmed the MBL behavior of the interacting GAA, a many-body counterpart of the intermediate phase has not been seen.

The experiments in Harvard, in particular~\textcite{Rispoli2019},  have concentrated on this critical regime, measuring observables like correlations between sites, entropy, and particle number fluctuations. The experimental procedure of~\textcite{Lukin2019,Rispoli2019} starts from a bosonic Mott insulator with one atom per site and measures the particle number per site after some evolution time. From this, different quantities can be computed. For instance, the probability $p_n$ of finding $n$ particles on a site determines the single-site von Neumann entropy, $S_{\rm vN}^{(1)} = - \sum_n p_n \log(p_n)$. For weak quasiperiodic potentials, the obtained value matches well with the expected thermal average, but it becomes reduced when the quasiperiodic modulation becomes stronger, approximately for $\lambda/t \gtrsim 4$, see Fig.~\ref{fig:MBL_exp}c, notably beyond the critical value of the single-particle AA model. In contrast, the entropy reaches a maximum within the transition regime as the system strongly fluctuates. Criticality is also characterized by strong $n$-site correlations $|G_c^{(n)}|$, shown in Fig.~\ref{fig:MBL_exp}(d). 

Bosonic systems have also been produced with up to 10 superconducting qubits, realizing Bose-Hubbard chains with an AA-like potential~\cite{Roushan2017}, and/or AA-modulated hopping~\cite{Li_2023}. All energy levels can be detected if prepared with only one or two bosons, revealing the fractal butterfly spectrum in the single-particle case. Using superconducting qubits, it has also been demonstrated that the level spacing of an interacting system is shifted towards smaller values by increasing the quasiperiodic modulation strength. By addressing bosonic systems, quantum simulators may finally help to solve the question of whether MBL of bosons exists~\cite{Choi2016}.

\subsection{Topological phenomena}
Topological quantum systems are characterized by a band structure with non-trivial topological properties, as in the case of topological insulators~\cite{Qi2011}, or by a topological many-body ground state, as in the case of FQH systems and topological superconductors~\cite{Nayak2008}. In all cases, the non-trivial topological properties are mathematically defined via non-local quantities, specifically non-zero Chern numbers, which are physically linked to robust properties, such as transport behavior or fractional quantum statistics of quasiparticles. Topological matter has been an intense research field in the past decades, mainly in the context of systems in ``regular" geometries and driven by the search for robust quantum systems. Quasicrystalline or fractal structures and higher-dimensional lattices provide interesting new aspects to the field of topological matter, as discussed in the following.

\subsubsection{Charge-pumping, Chern numbers, and gap labelling in 1D quasicrystals \label{subsec:topprop}}

There is a remarkable formal equivalence between AA-type 1D quasicrystals and the Harper-Hofstadter model~\cite{Harper1955,Hofstadter1976}. The latter describes electrons in a 2D periodic lattice subject to a uniform magnetic field along the $\hat{z}$ axis. In the Landau gauge, $\mathbf{A}=B(0,x,0)$, the Harper-Hofstadter Hamiltonian is diagonalized along the $y$-direction via a Fourier transform, resulting in an effective AA potential along the $x$-direction, with the phase parameter given by the wavevector $k_y$ and the modulation period by the magnetic flux. Hence, the magnetic field leads to a quasiperiodic structure when the lattice constant and magnetic length are incommensurate. This formal equivalence permits the use of topological concepts established for 2D crystals to be applied to 1D quasicrystals. In particular, as a paradigmatic model for the integer quantum Hall effect, the Harper-Hofstadter model has played a key role in connecting topological Chern numbers $C$ of the energy bands in the periodic system to quantized transport phenomena~\cite{Thouless1982,Fukui2005}. These integer numbers $C$ are obtained by considering a gapped band, that is, a set of states $|n(k_x,k_y)\rangle$ which are energetically separated from other states and which extend over the whole 2D Brillouin zone of the system, and integrating over the Berry curvature 
$\Omega(k_x,k_y)= \epsilon_{\mu\nu} \partial_\mu A_\nu(k_x,k_y)$ of this set of states:
%
%
\begin{align}
C= \frac{1}{2\pi i} \int_{\rm BZ} d^2 k\, \Omega(k_x,k_y).
\end{align}
%
%
Here, repeated indices are summed, $\epsilon_{\mu\nu}$ is the antisymmetric tensor, $\partial_\mu \equiv \frac{\partial}{\partial k_\mu}$, and the Berry connection $A_\mu(k_x,k_y)=\langle n(k_x,k_y) | \partial_\mu | n(k_x,k_y) \rangle$.
Physically, the Chern number quantifies the transverse conductivity of the band~\cite{Thouless1982}, as well as the number of states which, in the case of open boundary conditions, are localized at the system edge and connect different bands by winding through the energy gap~\cite{Hatsugai1993}. This relation between bulk transport properties and edge state is known as bulk-edge correspondence. With the Chern number being the band's topological property, the related physical behavior is robust against local perturbations. 

In the context of an AA model, Chern numbers are obtained by substituting $k_y$ with the phase angle $\phi$ in the quasiperiodic potential. Such a procedure can be applied to arbitrary 1D models with a cyclic parameter~\cite{Kraus2012,Kraus2012-2,Verbin2013,Grass2015}. Chern numbers of 1D quasicrystals have established a topological equivalence between different models, specifically between the AA model and a Fibonacci chain~\cite{Kraus2012-2,Verbin2013}. On the other hand, in the same way as tuning a magnetic field in a 2D lattice produces changes to the Hall conductivity, tuning the quasiperiodic modulation parameter in the AA model produces topological phase transitions in the 1D quasicrystal~\cite{Verbin2013}. The robustness of the Chern numbers against on-site interactions has been studied in~\textcite{Matsuda2014}.

\begin{figure*}[!t]
    \centering
    \includegraphics[width=0.99\textwidth]{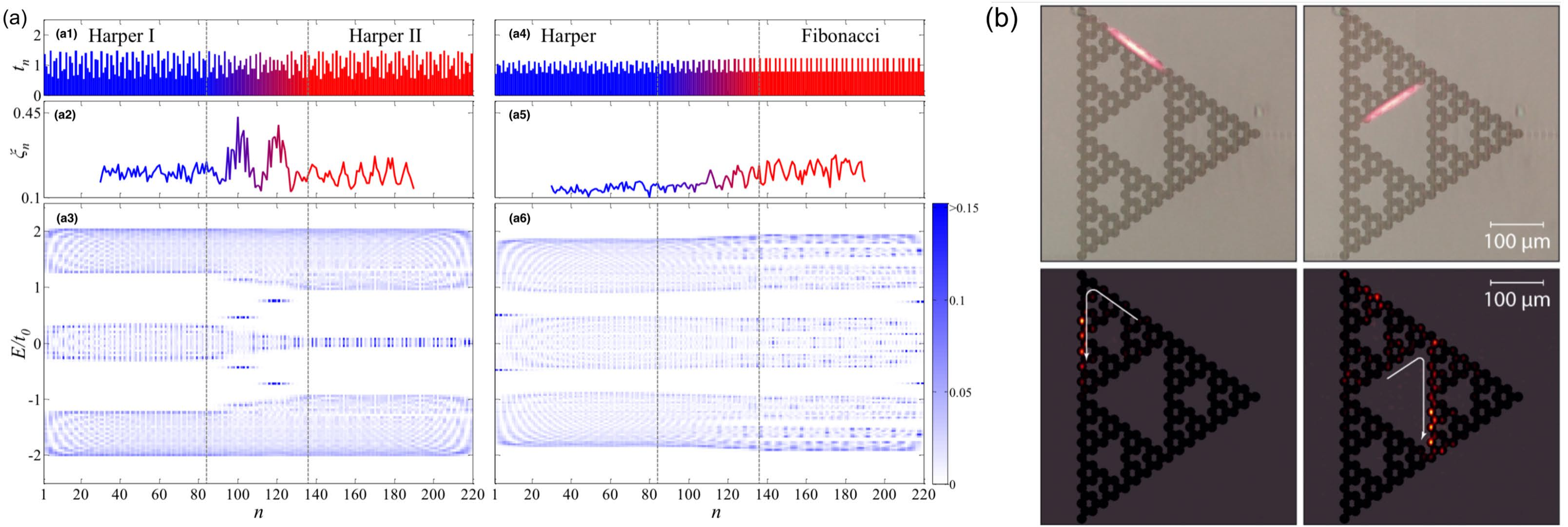}
    \caption{\textbf{Edge states in quasicrystals and fractal lattices.} (a) From~\textcite{Verbin2013}: Edge states at the interface of topologically different 1D quasicrystals. (a1) The Harper-Andre-Aubry quasicrystal is defined by quasiperiodic tunneling elements $t_n$ from the site $n$ to $n-1$, with different modulation parameters in the chain's right and left regions. Edge states localized at the interface are seen as return probability $\xi_n$ peaks in this region (a2) and as in-gap states in the energy spectrum (a3). In contrast, for an interface of a Harper-Andre-Aubry quasicrystal (with appropriate choice of modulation parameter) and a Fibonacci chain, no edge occurs, as shown in (a4-a6), indicating topological equivalence of the two quasicrystals. (b) From~\textcite{Biesenthal2022}: Exciting photons at inner or outer edges of a Sierpi{\'n}ski fractal (as shown in the upper figures), propagation along these edges is observed, as indicated by the output intensities after some propagation time, shown in the lower figures. The chirality of edge states is opposite on the inner and outer edges.
    \label{fig:topology}}
\end{figure*}

In the case of 1D quasicrystals, the physical relevance of non-zero Chern numbers is evidenced by pumping experiments, where a quantized amount of ``charge" is pumped through the chain when the cyclic Hamiltonian parameter is adiabatically changed. The amount of the pumped ``charge" is defined by the Chern numbers of the filled bands. Such topological charge pumping has already been proposed by~\textcite{Thouless1983} in the context of 1D periodic (super)lattices. It has been realized in different photonic quasicrystals built from coupled waveguides in~\textcite{Kraus2012,Verbin2013} by adiabatically sweeping the phase of the quasiperiodic potential along the propagation axis. This setup has also served to visualize the presence or absence of localized states at the boundary between different quasicrystals~\cite{Verbin2013}, depending on whether the quasicrystals are topologically equivalent or not, see Fig.~\ref{fig:topology}a. With cold atoms, one can fill the lowest band either as a fermionic band insulator~\cite{Nakajima2016} or a bosonic Mott insulator~\cite{Lohse2016} and obtain a quantized charge transport of one site per modulation cycle.

From a static point of view, that is, without change of a Hamiltonian parameter, the topology of 1D quasicrystals is captured by the gap labeling theorem~\cite{Bellissard1989,Bellissard1992}. This theorem generally applies to 1D Schr\"odinger equation with bounded potential, stating that within the energy gaps, the integrated density of states $\rho_{\rm int}(E) = \int_{-\infty}^{E} \rho(E') dE'$ takes quantized values expressed by integer labels. For instance, for a Fibonacci chain, $\rho_{\rm int}(E)$ is given by
%
%
\begin{align}
\rho_{\rm int}(E) = P + \tau^{-1} Q,
\end{align}
%
%
where $\tau=(1+\sqrt{5})/2$ is the golden mean, and $P,Q$ are integers. Like the Chern numbers, these gap labels $P$ and $Q$ are topologically robust: a perturbation does not change them unless the perturbation closes the band gap. The accuracy of the gap labeling theorem in Fibonacci chains has been demonstrated experimentally using cavity-polaritons~\cite{Tanese2014,Baboux2017}. These observations suggest that Chern numbers and gap labels are two sides of the same coin~\cite{Dareau2017}.
%
%
%
%

\subsubsection{ 4D quantum Hall effect in 2D quasicrystals \label{Sec:QH4D} }
The mapping between 1D superlattices and 2D quantum Hall models can be extended to higher dimensions: Modulating a 2D lattice along both directions, it can be mapped onto a 4D quantum Hall system~\cite{Kraus2013}. In such a system, a Hall current $I_{\alpha}$ is induced by the combination of a perpendicular electric field perturbation $E_{\beta}$ and a magnetic field perturbation $B_{\gamma\delta}$ in the $\gamma\delta$-plane perpendicular to $\alpha$ and $\beta$. 
Such a response is described by novel topological concepts, specifically the second Chern number. It can be obtained by decomposing the (2+2)-dimensional system into two (1+1)-dimensional systems along each physical dimension $\alpha=x,y$. The Berry curvature $\Omega_\alpha(k_\alpha,\phi_\alpha)$ of the (1+1)-dimensional system along $\alpha$ is defined through the wavevector $k_\alpha$ and the modulation $\phi_\alpha$. The second Chern number is then obtained by integrating the product of Berry curvatures over the 4D Brillouin zone~\cite{Qi_2008,Mochol-Grzelak_2019},
%
%
\begin{align}
    C_2 = \frac{1}{4\pi^2} \int_{\rm BZ} \Omega_x \Omega_y dk_x dk_y d\phi_x d\phi_y.
\end{align}
%
%
If the bands are factorizable along $x$ and $y$, the second Chern number can be decomposed into products of the first Chern numbers. This decomposition then also applies to the edge states: These states can either be 0D corner states, obtained as a product of the 0D edge states of the individual 1D systems, or a 1D state, obtained as the product of a 0D edge state and a 1D bulk state.

Experimentally, these different aspects of the 4D quantum Hall effect have been probed with atomic and photonic quantum simulators~\cite{Lohse2018,Zilberberg2018}. Specifically, 2D photonic waveguides with modulated coupling along the propagation direction have probed 4D quantum Hall response via photon pumping from edge to edge and corner to corner~\cite{Zilberberg2018}, see also Fig.~\ref{fig:4D_Photonic}.
Cold atoms in an optical 2D superlattice have been used to study the bulk behavior of a 4D quantum Hall system~\cite{Lohse2018}. This setup has allowed for measuring the second Chern number via the response of the atoms to a perturbing electric field, mimicked by a phase shift of the superlattice potential, and to a magnetic perturbation, mimicked by a small tilt of the long lattice with respect to the corresponding short lattice.

We also note the recent implementation of the 4D quantum Hall effect in a system with synthetic extra-dimensions encoded in the large electronic spin manifold of dysprosium atoms~\cite{Bouhiron2022}.

\subsubsection{Topological classification}

The ten-fold topological classification~\cite{Ryu2010} considers the behavior of a Hamiltonian under three fundamental discrete symmetries: time-reversal~$\mathcal{T}$, charge-conjugation $\mathcal{C}$, and the product of the two $\mathcal{S}= \mathcal{T}\cdot\mathcal{C}$.  The ten-fold way establishes that a system that breaks all these symmetries can have non-zero Chern numbers only in even dimensions. The purported contradiction of this scheme through 1D quasicrystals is resolved by noting that the Chern numbers for the quasicrystals do not classify a single 1D Hamiltonian, but a parametrized family of models~\cite{Madsen2013}. From this perspective, the ten-fold topological classification remains unchanged for quasicrystals.

Beyond the ten-fold way, there are also crystal symmetries that can protect topological phases. This can give rise to weak topological insulators/superconductors or higher-order topological insulators (HOTIs)~\cite{Chiu2016}. 
In total, there are 528 (1651) magnetic space group symmetries in 2D (3D), which can be used for classifying topological quantum matter by crystal symmetries~\cite{Watanabe2018}.
However, since quasicrystals, by definition, support symmetries that cannot exist in crystals, it is not surprising that topological phases without crystalline counterparts become possible in quasiperiodic systems~\cite{Varjas2019,Chen_2020,Else2021,Spurrier_2020}.
In~\textcite{Varjas2019}, it has been shown that a HOTI with gapless topological corner modes exists for a 2D quasicrystal with eightfold rotational symmetry. A classification scheme for topological phases in quasicrystals has been presented in~\textcite{Else2021}. Still,
the topological richness of quasicrystals has not yet been fully explored. HOTIs can also occur in higher-dimensional spaces. In~\textcite{Dutt2020}, a 4D HOTI has been realized using the synthetic frequency dimension of photonic molecules.

\subsubsection{Topological edge states in fractals and quasicrystals}

New topological phases are also expected or already observed in fractal and hyperbolic lattices. Whereas for standard geometries, the bulk-edge correspondence establishes an important relation between topological bulk behavior (Chern numbers) and topological edge states, extended bulk regions are lacking in fractal systems. Non-interacting topological systems (i.e., topological insulators and superconductors) on Sierpi{\'n}ski fractals have been studied theoretically in~\textcite{Brzezinska2018,Pai2019,Iliasov2020,Fremling2020} \textcolor{blue}{, and in the Penrose tiling~\cite{Grossi_e_Fonseca_2023}}. In these systems, the Hall conductivity is neither quantized nor proportional to the Chern number of the filled bands. However, splitting into various sub-bands makes it difficult to discern whether a band is filled. The fate of edge states is particularly intriguing since, on a fractal lattice, most sites are near some edge of the system. Topological edge states in a Sierpi{\'n}ski gasket have recently been observed in a photonic setup~\cite{Biesenthal2022,Li2023}, see Fig.~\ref{fig:topology}(b).

Anyons are the hallmark of interacting topological systems. The question of how these exotic quasiparticles are affected by fractal dimensions has been theoretically addressed in~\textcite{Manna2020,Manna2022}, giving exact parent Hamiltonians for analytic FQH  wave functions on Sierpi{\'n}ski fractals and demonstrating anyonic braiding behavior of their quasiparticles. The entanglement entropy scaling in fractional Chern states on fractal lattices has been studied in~\textcite{Li2022}, and a violation of the area-law indicates that entanglement entropy is not suited to characterize topological phases on fractal lattices.

Theoretically simpler than the FQH effect are topological superconductors, such as the paradigmatic Kitaev chain~\cite{Kitaev2001}. Even on a mean-field level, these systems support non-Abelian anyons, specifically Majorana zero modes localized at the ends of the chain. These quasiparticles, which are their own antiparticles, possess intriguing topological properties that can be used to encode and process quantum information.
Signature of Majorana modes have been observed in chains of magnetic atoms placed on top of superconductors~\cite{NadjPerge_2014,Schneider2022,Yazdani_2023}; however, there is no consensus on the topological origin of the observed boundary modes~\cite{Kuester_2022}. Following the work by~\textcite{Soldini_2023}, where 2D regular lattices of magnetic impurities were created on a superconductor, we envision that this technique could be extended to construct quasicrystal and fractal lattices of magnetic impurities to realize topological superconductors in non-standard geometries.
Quasiperiodic models, and especially the AA model, have attracted a lot of attention in the context of topological superconductivity~\cite{Cai2013,DeGottardi2013,DeGottardi2013-2,Zeng2016,Wang2016,Yahyavi2019,Fraxanet2021}. The quasiperiodic potential provides an excellent testing ground for the robustness of the topological phase, as it does not break the protecting symmetries. Both analytic and numerical studies have explicitly confirmed that the topological phase and the Majorana quasiparticle persist in the presence of a weak quasiperiodic potential, but a strong incommensurate modulation drives a localization transition that destroys the topological phase. Notably, in some cases, quasiperiodic modulation of the chemical potential~\cite{Fraxanet2021} or of the hopping~\cite{Wang2016} might even stabilize the topological phase.

The recent realization of hyperbolic lattices has inspired the development of hyperbolic lattice crystallography~\cite{Boettcher2022}, including topological phases such as hyperbolic Haldane and Kane-Mele models~\cite{Urwyler2022, Zhang_2022}. A relevant feature for topological phases is the extensive scaling of the boundary for hyperbolic lattices, where a macroscopic fraction of all states contributes to topological edge states. Hyperbolic lattices can feature non-trivial second Chern numbers for zero first Chern numbers~\cite{Zhang2023} as well as higher-order topology with an unconventional number of corner modes, not allowed in crystalline materials~\cite{Tao2023}. A hyperbolic photonic topological insulator was recently realized using coupled ring resonators on silicon chips~\cite{Huang2024}.

\subsection{Cosmological models}
The engineering of systems with curved geometries provides a new tool for studying gravitational physics and cosmology in the laboratory. As examples, we first consider the quantum simulation of gravitational effects in lattices, which explicitly implement curvature through the appropriate connectivity design. We then also mention the wide field of analog gravity: The analogy between sound propagation on a background hydrodynamic flow and field propagation in curved spacetime, first pointed out by~\textcite{Unruh1981}, allows for studying gravitational phenomena even in flat systems. A similar analogy can also be made between light propagation in inhomogeneous media and field propagation in curved spacetime~\cite{Khveshchenko_2015}.

%
%
\begin{figure*}
    \includegraphics[width=\textwidth]{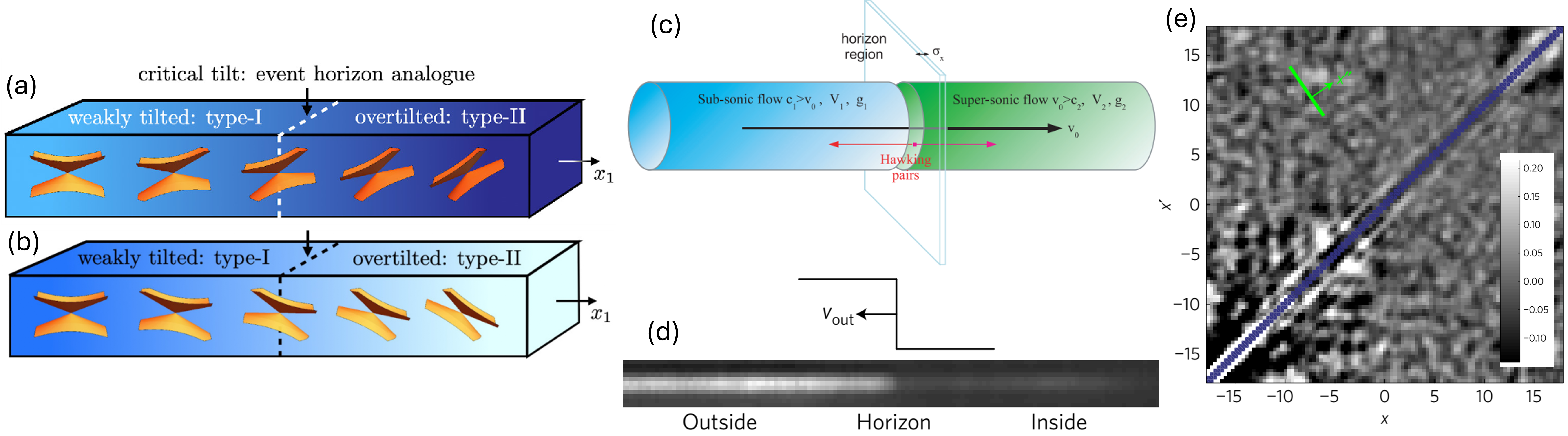}
    \caption{\textbf{Simulations of cosmological models.} (a) Sketch of a Weyl semimetal with tilt varying along $x_1$ realizing a white hole (a) and a black hole analog (b), respectively.  Adapted from~\textcite{Sabsovich2022}. (c) Sketch of a ``sonic horizon" formed at the transition between subsonic and supersonic flow in a quantum fluid (either an atomic BEC or a polariton BEC). Adapted from~\textcite{Carusotto_2008}. (d) The experimental realization of a "sonic horizon" within a 1D BEC of 87$^\text{Rb}$ atoms is presented. The horizon is created by moving a potential step across the condensate from left to right at a speed of $v_\text{out}=0.24$ mm/s. In the reference frame of the potential, this is equivalent to the arrangement in (c): a stationary horizon and a condensate flowing from left to right at the speed $v_0=v_\text{out}$.  (e) Experimental results of the second-order correlation function  $G^{(2)}(x,x')$, as defined by Eq.~\eqref{eq:G2}. The horizon is located at $x=0$. The two dark bands emanating from the horizon signify the correlated pairs of phonons, one moving inward and the other moving outward from the horizon. Adapted from~\textcite{Steinhauer2016}.
    \label{fig_weyl}}
\end{figure*}
%
%
The Unruh effect~\cite{Fulling_1973,Davies_1975,Unruh_1976} and the Hawking radiation~\cite{Hawking_1974} are two important aspects related to the physics of black holes. Both phenomena are based on predictions of quantum field theory in curved spacetimes and non-inertial reference frames. 
The Unruh effect predicts that an observer with constant acceleration $a$ through empty space will perceive a thermal bath characterized by the Unruh temperature $k_\text{B} T_\text{U}=\hbar a/2\pi c$, where $c$ is the speed of light in vacuum, $\hbar$ is the reduced Planck constant, and $k_\text{B}$ is the Boltzmann constant. The same is not true for an inertial observer. 
Hawking predicted that black holes are far from being emission-free, and they should emit a steady flux of thermal radiation, known as Hawking radiation, with a temperature proportional to $\kappa$, the gravitational field strength at the event horizon $k_\text{B} T_\text{H}=\hbar \kappa/2\pi c$. Both effects are extremely weak and have not been observed using astronomical techniques. In this respect, analog quantum simulators of these are very appealing.

For the case of the Unruh effect,~\textcite{Rodriguez-Laguna2017,Kosior2018} proposed an implementation with cold atoms in an optical lattice. The general idea is to simulate a tuneable Dirac-Hamiltonian system that can interpolate between a Minkowski (rest frame) and a Rindler (constant accelerated frame) configuration. Specifically, the key difference between the two implementations is the gauge field characterizing the cold atoms' hopping terms in the optical lattice. These two different Hamiltonians are implemented, \emph{e.g.}, in a square lattice with constant hopping term $J$ along the two directions and by adding a gradient diagonally to the lattice. The difference in the hopping terms between the Minkowski and the Rindler implementation can be obtained by applying an additional Raman laser collinear with the lattice beams. Finally, information about the Unruh effect is obtained by measuring the Wightman response, which is a two-point correlation function~\cite{Rodriguez-Laguna2017}. A similar implementation for interacting cold atoms is proposed in~\textcite{Kosior2018}.

There is also an analogy between the general relativity and the Hamiltonian of Weyl's systems~\cite{Volovik_2016}. A Weyl Hamiltonian can be obtained as low-energy excitations of Weyl semimetals~\cite{Armitage_2018}. The Hamiltonian describing Weyl semimetals reads
%
%
\begin{equation}\label{Ham_Weyl}
    \mathcal{H}_\text{Weyl}=\pm v_\text{F}\bm{\sigma}\bm{p}+\mathbb{I}_2\bm{V}_t \bm{p}
\end{equation}
%
%
where $v_\text{F}>0$ is the Fermi velocity, $\bm{p}$ the three-dimensional momentum, and $\bm{\sigma}$ the vector of Pauli matrices. For small tilts, $|\bm{V}_t| < v_\text{F}$, the Weyl cone is denoted as type I. For large tilts $|\bm{V}_t| > v_\text{F}$, the Weyl cone is overtilted and called type II~\cite{Armitage_2018}. Type I Weyl cones correspond to light cones in a flat Minkowski spacetime, whereas type II Weyl nodes to light cones that tilt close to black or white holes. The former is for tilting towards the horizon, and the latter is for tilting away from the horizon | see Fig.~\ref{fig_weyl}(a-b). There is a simple mapping between Eq.~\eqref{Ham_Weyl} and the spacetime metrics:  one must define the frame fields $e_\mu^i$ and their inverse $\underline{e}^\nu_j$~\cite{Volovik_2016,Volovik_2017}. In general relativity, a frame field defines a local orthonormal coordinate system at each point in space-time. They are connected to the spacetime metric as $g_{ij}=\eta_{\mu\nu}\underline{e}^\mu_i\underline{e}^\nu_j$, where $\eta_{\mu\nu}$ represents the Minkowski metric. The terms in Eq.~\eqref{Ham_Weyl} can be rearranged to introduce frame fields as the tensor connecting Pauli matrices and momenta; in this way, we obtain:
%
%
\begin{equation}\label{Ham_Weyl_gen}
    \mathcal{H}_\text{Weyl}^\text{gen}=-\text{i} v_\text{F}\sigma^\mu e_\mu^j\left(\partial_j+\frac{1}{2}\underline{e}_\mu^j(\partial_k e_\nu^k)\right),
\end{equation}
%
%
with $\sigma^0=\mathbb{I}_2$. In Eq.~\eqref{Ham_Weyl_gen}, Roman (Greek) indices run over spatial (space-time) coordinates, and we are assuming the summation over identical indices. The essence of the mapping between Weyl Hamiltonians and black holes is that the frame fields $e_\mu^i$ defined from Eq.~\eqref{Ham_Weyl_gen} coincide with the ones of the Gullstrand-Painlev\'e-Schwarzschild metric associated with $ds^2=c^2dt^2-(d\bm{r}-\bm{V}(\bm{r})dt)^2$,
with the tilt $\bm{V}_t /v_\text{F}$ taking the role of the velocity $\bm{V}/c$. Thus, a Weyl Hamiltonian with a continuous tilt changing from $\bm{V}_t (\bm{r}) = 0$ to $|\bm{V}_t(r)| > v_\text{F}$ can be mapped to the metric of a black or white hole in Gullstrand-Painlev\'e coordinates. The location of the critical tilt, $|\bm{V}_t(r)| = v_\text{F}$, is the analog of the event horizon.

In~\textcite{De_Beule2021,Sabsovich2022}, a continuous smooth junction between a type I and type II Weyl semimetal has been proposed to study an artificial event horizon. Transport experiments can give rise to the signature of Hawking-like radiation~\cite{De_Beule2021}. When considering a more realistic implementation of Weyl semimetals in a lattice system, there is a doubling of the Weyl cone. Consequently, this gives rise to new effects dubbed Hawking fragmentation and Hawking attenuation of a wave packet approaching the event horizon~\cite{Sabsovich2022}. Junctions of this type cannot be implemented in realistic materials hosting Weyl semimetals but could be engineered in acoustic crystals~\cite{Yang_2016,Peri_2019}, photonics~\cite{Huang_2020}, topoelectrical circuits~\cite{Rafi_Ul_Islam_2020}, and cold atoms~\cite{Xu_2016}. A recent proposal by~\textcite{Stalhammar_2023} extends these results to the realm on non-Hermitian $\mathcal{PT}$ symmetric systems.

The exceptional tunability of certain quantum systems, notably cold atomic systems and quantum fluid of light (i.e., non-linear transport of polariton BEC), renders them ideal for simulating particle flows that mimic curved spacetime for the system's phononic excitations. For instance, the concept of ``sonic black holes" within a  BEC was proposed~\cite{Garay2000,Balbinot2008,Carusotto2013}, leading to the realization of a polaritonic black hole~\cite{Nguyen2015} and the subsequent atomic realization of sonic black holes as experimental tools for investigating Hawking radiation~\cite{Steinhauer2016,Munoz2019,Kolobov2021}. The main idea behind these realizations is similar to the Unruh effect in classical fluids but now applied to quantum fluids: the ``sonic horizon" separates a subsonic upstream region ($v_0<c_1$) from a supersonic downstream region ($v_0>c_2$), as illustrated in Figs.~\ref{fig_weyl}(c,d), where $v_0$ is the speed of the BEC flow and $c_{1,2}$ are the speed of sound in the upstream and downstream regions, respectively. To probe the Hawking radiation in the case of 1D black holes, in~\textcite{Balbinot2008,Carusotto_2008} has been proposed to use the second-order correlation function of density fluctuation, given by:
%
%
\begin{equation}\label{eq:G2}
G^{(2)}(x, x') = \frac{\langle :\!n(x) n(x')\!:\rangle}{\langle n(x) \rangle \langle n(x') \rangle}
\end{equation}
%
%
with $:\!\square\!:$ denoting normal ordering.
The signature of Hawking radiations, i.e., correlated phonon pairs that are continuously created at the horizon, is a symmetric pair of negative correlation stripes in the mapping $G^{(2)}(x, x')$ with a slope of $(v_0-c_2)/(v_0-c_1)$. This corresponds to the propagation of correlated phonon pairs in the upstream and downstream region at speeds $v_0-c_1 < 0$ and $v_0-c_2 > 0$, respectively. This correlation has been numerically demonstrated using Monte Carlo simulations~\cite{Carusotto_2008,Nguyen2015} and experimentally observed in atomic BEC system~\cite{Steinhauer2016}, as shown in Fig.~\ref{fig_weyl}(e).  Moreover, the ability to modify the curved metric for phonons in a BEC dynamically has facilitated the exploration of cosmological models, providing insights into the equations describing the universe's expansion dynamics~\cite{Viermann2022}.

\section{Conclusion and Outlook }
This colloquium has aimed at reviewing the state-of-the-art quantum engineering techniques and their reachable physical phenomena from a context that concentrates on exotic geometries. To this aim, we have covered various systems (atomic, electronic, and photonic ones), highlighting unique opportunities offered by different platforms and introducing general concepts such as Floquet engineering and synthetic dimensions. 
Many of these developments involve quantum control on the microscopic level and, therefore, resonate with strong efforts of building programmable quantum simulators. Such devices may then serve computational purposes and eventually demonstrate a quantum computational advantage, e.g. via boson sampling using a network of photons~\cite{Zhong_2020}, or via quantum optimization using an Rydberg atoms in optical tweezer arrays~\cite{Ebadi_2022}.

In this colloquium, we have concentrated on specific physical implications of the exotic geometries. This includes (i) localization phenomena, (ii) topological phenomena, (iii) analog black holes, and cosmology. These aspects already contain a rich variety of physical behavior on the single-body level, such as mobility edges in low dimensions (smaller than 3) or integer quantum Hall effect in high dimensions (larger than 3). While some aspects, such as topological classification in non-periodic crystals, remain an outstanding challenge, many features related to single-particle physics in exotic lattices have been theoretically understood and experimentally realized. Much less explored, though, is the complicated subject of quantum many-body physics in such geometries. In this regard, the present colloquium article has provided a brief perspective on many-body localization. Theoretical research on this topic has indeed been ignited by the possibility of experimentally building synthetic quasicrystals that show this intriguing behavior or similar behavior (such as glassy phases). Yet there is open debate on the true existence of MBL phase, particularly in more than one dimension.
Very little is known regarding topological many-body phenomena in quasicrystals or fractals. Still, the rich single-particle physics triggers hope for an even richer many-body scenario. Given the complexities in simulating such systems classically, the quantum simulators may develop their full potential of opening the door to deep questions of quantum many-body physics. With the engineering of exotic kinetic terms being achieved now, the basis has been set for future explorations of the interacting world of fractals, quasicrystals, and curved and higher-dimensional spaces. 

While this review has focused on quantum simulators where the setup imposes exotic geometries, there are also fascinating opportunities to observe similar geometric structures emerging spontaneously, for instance, due to multi-body interactions. An example are fracton phases, where fractal structures arise as the low-dimensional subspace of exotic quasi-particles with restricted mobility~\cite{Nandkishore_2019,Pretko_2020}. While the requirement of multispin interactions is difficult to engineer, Rydberg atoms have recently been proposed as a suitable experimental platform~\cite{Myerson-Jain2022}. This might open another intriguing avenue towards quantum systems with exotic geometries~\cite{Vijay_2016}.

\begin{acknowledgments} 
T.G. and D.B. acknowledge fruitful discussions with Geza Giedke.
D.B. acknowledges the support from the Spanish MICINN-AEI 
through Project No.~PID2020-120614GB-I00~(ENACT), the Transnational Common Laboratory $Quantum-ChemPhys$, and the Department of Education of the Basque Government through
the project PIBA\_2023\_1\_0007 (STRAINER). T.G. and D.B. acknowledge the financial support received from the IKUR Strategy under the collaboration agreement 
between the Ikerbasque Foundation and DIPC on behalf of the Department of Education of the 
Basque Government, and by the Gipuzkoa Provincial Council within the QUAN-000021-01 project. 
T.G. acknowledges funding by the Department of Education of the Basque Government through the project PIBA\_2023\_1\_0021 (TENINT), by the Agencia Estatal de Investigación (AEI) through Proyectos de Generación de Conocimiento PID2022-142308NA-I00 (EXQUSMI), and that this work has been produced with the support of a 2023 Leonardo Grant for Researchers in Physics, BBVA Foundation. The BBVA Foundation is not responsible for the opinions, comments, and contents included in the project and/or the results derived therefrom, which are the total and absolute responsibility of the authors. H.S.N. is funded by the French National Research Agency (ANR) under the project README (ANR-22-CE09-0036-01). The work of C.W. is funded by the Cluster of Excellence “CUI: Advanced Imaging of Matter” of the Deutsche Forschungsgemeinschaft (DFG)—EXC 2056 —Project ID No. 390715994 and by the European Research Council (ERC) under the European Union’s Horizon 2020 research and innovation program under Grant Agreement No. 802701. U.B.
is also grateful for the financial support of the IBM Quantum Researcher Program.
M.L. acknowledges ERC AdG NOQIA; MCIN/AEI (PGC2018-0910.13039/501100011033, CEX2019-000910-S/10.13039/501100011033, Plan National FIDEUA PID2019-106901GB-I00, Plan National STAMEENA PID2022-139099NB-I00 project funded by MCIN/AEI/10.13039/501100011033 and by the ``European Union NextGenerationEU/PRTR'' (PRTR-C17.I1), FPI); QUANTERA MAQS PCI2019-111828-2); QUANTERA DYNAMITE PCI2022-132919 (QuantERA II Programme co-funded by European Union's Horizon 2020 program under Grant Agreement No. 101017733), Ministry of Economic Affairs and Digital Transformation of the Spanish Government through the QUANTUM ENIA project call -- Quantum Spain project, and by the European Union through the Recovery, Transformation, and Resilience Plan -- NextGenerationEU within the framework of the Digital Spain 2026 Agenda; Fundaci{\' o} Cellex; Fundaci{\' o} Mir-Puig; Generalitat de Catalunya (European Social Fund FEDER and CERCA program, AGAUR Grant No. 2021 SGR 01452, QuantumCAT \textbackslash U16-011424, co-funded by ERDF Operational Program of Catalonia 2014-2020); Barcelona Supercomputing Center MareNostrum (FI-2023-1-0013); EU Quantum Flagship (PASQuanS2.1, 101113690); EU Horizon 2020 FET-OPEN OPTOlogic (Grant No. 899794); EU Horizon Europe Program (Grant Agreement 101080086 -- NeQST), ICFO Internal ``QuantumGaudi'' project; European Union's Horizon 2020 program under the Marie-Sklodowska-Curie grant agreement No. 847648; ``La Caixa'' Junior Leaders fellowships, ``La Caixa'' Foundation (ID 100010434): CF/BQ/PR23/11980043. Views and opinions expressed are, however, those of the author(s) only and do not necessarily reflect those of the European Union, European Commission, European Climate, Infrastructure and Environment Executive Agency (CINEA), or any other granting authority. Neither the European Union nor any granting authority can be held responsible for them.
\end{acknowledgments}

\bibliography{bib}

\end{document}